\documentclass[prx,twocolumn,english,superscriptaddress,floatfix,longbibliography]{revtex4-2} 

\usepackage{enumerate,amsmath}
\usepackage{hyperref,graphicx}
\usepackage{color}
\graphicspath{{./Figures}}
\usepackage{here}
\usepackage{xcolor}
\hypersetup{
    colorlinks=true,
    citecolor=blue,
    linkcolor=magenta,
    urlcolor=magenta,
}

\usepackage{braket}
\usepackage{comment}
\usepackage{algorithmic}
\usepackage{algorithm}
\usepackage{amsmath}
\usepackage{amsmath,amssymb}
\usepackage{amsfonts}
\usepackage{overpic}
\usepackage{latexsym}
\usepackage{amsmath,amssymb}
\usepackage{braket}
\usepackage{physics}
\usepackage{quantikz}
\usepackage{amsmath}
\usepackage{longtable}
\usepackage{tabularx}    
\usepackage{array}       
\usepackage{booktabs}   
\usepackage{array}      
\usepackage{amsmath}    
\usepackage{xcolor}     
\usepackage{siunitx}    
\usepackage{comment}
\usepackage{graphicx,adjustbox}
\usepackage[caption=false]{subfig}
\graphicspath{{./Figure/}}

\makeatletter
\renewcommand*{\@opargbegintheorem}[3]{\trivlist
      \item[\hskip \labelsep{\bfseries #1\ #2}] \textbf{(#3)}\ \itshape}
\makeatother

\begin{document}

\title{Fisher information based lower bounds on the cost of quantum phase estimation}

\author{Ryosuke Kimura}
\email{u294831i@ecs.osaka-u.ac.jp}
\affiliation{Graduate School of Engineering Science, Osaka University, 1-3 Machikaneyama, Toyonaka, Osaka 560-8531, Japan}

\author{Kosuke Mitarai}
\email{mitarai.kosuke.es@osaka-u.ac.jp}
\affiliation{Graduate School of Engineering Science, Osaka University, 1-3 Machikaneyama, Toyonaka, Osaka 560-8531, Japan}
\affiliation{Center for Quantum Information and Quantum Biology, The University of Osaka, 1-2 Machikaneyama, Toyonaka 560-0043, Japan}

\begin{abstract}
Quantum phase estimation (QPE) is a cornerstone of quantum algorithms designed to estimate the eigenvalues of a unitary operator. QPE is typically implemented through two paradigms with distinct circuit structures: quantum Fourier transform-based QPE (QFT-QPE) and Hadamard test-based QPE (HT-QPE). Existing performance assessments fail to separate the statistical information inherent in the quantum circuit from the efficiency of classical post-processing, thereby obscuring the limits intrinsic to the circuit structure itself. In this study, we employ the Fisher information and the Cramér-Rao lower bound to formulate the performance limits of circuit designs independent of the efficiency of classical post-processing. Defining the circuit depth as $T$ and the total runtime as $t_{\text{total}}$, our results demonstrate that the achievable scaling is in fact constrained by a non-trivial lower bound on their product $Tt_{\text{total}}$, although previous studies have typically treated the circuit depth $T$ and the total runtime $t_{\text{total}}$ as separate resources. Notably, QFT-QPE possesses a more favorable scaling with respect to the overlap between the input state and the target eigenstate corresponding to the desired eigenvalue than HT-QPE. Numerical simulations confirm these theoretical findings, demonstrating a clear performance crossover between the two paradigms depending on the overlap. Furthermore, we verify that practical algorithms—specifically the quantum multiple eigenvalue Gaussian filtered search (QMEGS) and curve-fitted QPE—achieve performance levels closely approaching our derived limits. By elucidating the performance limits inherent in quantum circuit structures, this work concludes that the optimal choice of circuit configuration depends significantly on the overlap.
\end{abstract}

\maketitle

\section{Introduction}

Quantum phase estimation (QPE) is one of the most useful subroutines in quantum computing and plays a crucial role in many promising applications such as quantum chemistry and material science~\cite{aspuru2005simulated, whitfield2011simulation, reiher2017elucidating}. Given a unitary matrix $U$ and an input state $\ket{\psi}$, the task of QPE is to estimate a phase $\theta$ of an eigenvalue of $U$ within a specified precision. This estimation is achieved by repeatedly applying controlled-$U$ operations, which serve as the computational cost in QPE algorithms. To quantify the computational costs required for this task, researchers have focused on two key metrics. The first is the maximum number $T$ of applications of controlled-$U$ that represents the circuit depth. The second is the total number $t_{\text{total}}$ of applications of controlled-$U$ that represents the total runtime.

In the literature, two representative paradigms have emerged to implement these estimations. The first is the quantum Fourier transform-based QPE (QFT-QPE)~\cite{nielsen2010quantum, 2409.15752, Kaiser, tapering2024optimal, qftqpe_fisher, quantum_algorithms_revisited}, which utilizes an ancilla register to perform a collective measurement of the phase information. Although it provides a high-precision estimate in a single run, the requirement for numerous ancillary qubits makes it more suitable for fault-tolerant quantum computer(FTQC). The second is the Hadamard-test-based QPE (HT-QPE)~\cite{Lin_QPE, Dong_QPE, Wang_QPE, 2402.01013, QCELS, CS,RPE}, a more recent and resource-efficient alternative that has become central to the study of early-FTQC. HT-QPE reduces the hardware overhead to a single ancilla by decomposing the QFT-QPE's coherent operations into multiple, simpler circuits and classical statistical processing.

In general, any QPE protocol comprises two distinct degrees of freedom: the first is the quantum circuit design that determines the measurement probability distribution, and the second is the classical processing algorithm used to process the measurement outcomes.
Much of the existing literature, however, analyzes these two components as an inseparable unit.
Consequently, it remains unclear whether a specific performance limitation stems from the bounds of the quantum circuit itself or from the inefficiency of the classical processing.
Previous studies have performed comparisons of the parameter $T$ and $t_{\text{total}}$ required to achieve a target precision $\varepsilon$ through theoretical analyses and numerical simulations, fixing specific pairs of quantum circuit configurations and classical processing algorithms~\cite{2402.01013, QCELS, MM-QCELS, RPE, Lin_QPE}.
In particular, in comparisons based on numerical simulations, it has been reported that QFT-QPE is inferior to HT-QPE in terms of $T$~\cite{2402.01013, QCELS, MM-QCELS, RPE}.
However, since the classical processing algorithms applied to QFT-QPE in these comparisons were not necessarily optimized, it is uncertain whether the reported performance gaps truly indicate the inferiority of QFT-QPE.

In this work, we provide a theoretical framework based on Fisher information to clarify the cost lower bounds of QPE and evaluate the efficiency of practical classical estimators. By using the Cramér-Rao lower bound, we derive lower bounds for various schemes on the cost product $Tt_{\text{total}}$. This approach allows us to identify the performance limits inherent in the circuit design itself, independently of the classical algorithm used. While previous studies have typically treated $T$ and $t_{\text{total}}$ as separate resources, our results show that the achievable scaling is in fact constrained by a nontrivial relation on the product $T \, t_{\text{total}}$. Our analysis reveals that the costs of the two representative schemes exhibit distinct scaling behaviors with respect to the overlap $c_i$ between an input state and the target eigenstate. Specifically, we show that QFT-QPE satisfies $T t_{\text{total}} = \Omega(c_i^{-1} \text{MSE}_i^{-1})$, whereas HT-QPE satisfies $T t_{\text{total}} = \Omega(c_i^{-2} \text{MSE}_i^{-1})$. This difference implies a crossover in their efficiency; within the range of parameters investigated in our numerical analysis, we observe that HT-QPE exhibits a smaller lower bound in the high-overlap regime, while QFT-QPE is more favorable in the low-overlap regime. We further examine the constant factors appearing in these lower bounds and find that, among representative HT-QPE protocols, the lower bounds are of comparable magnitude.

Finally, we compare these derived lower bounds with the actual costs that can be achieved with representative QPE algorithms through numerical simulations. We demonstrate that for schemes such as quantum multiple eigenvalue Gaussian filtered search, known as QMEGS~\cite{2402.01013}, and curve-fitted QPE~\cite{curve_qpe}, the product $T t_{\text{total}}$ closely approaches our derived lower bounds, whereas other HT-QPE variants exhibit a noticeable gap. This comparison reveals that, although the achievable circuit-level performance of HT-QPE protocols is similar up to constant factors, the overall performance depends on the efficiency of the classical post-processing, with QMEGS providing the most efficient realization in the asymptotic regime.

The remainder of this paper is organized as follows. In Sec.~\ref{Preliminary}, we describe our phase estimation setting, introduce the Fisher information and the Cramér-Rao lower bound, and summarize the QFT-QPE and HT-QPE. In Sec.~\ref{Theory}, we derive a lower bound on the cost product $Tt_{\text{total}}$ for our non-adaptive phase estimation scheme using the Fisher information matrix and the Cramér-Rao lower bound. In Sec.~\ref{main_results}, we instantiate this bound for QFT-QPE and HT-QPE, analyze the resulting asymptotic scaling and crossover behavior, and compare the costs lower bound with the actual costs of practical algorithms. Finally, Sec.~\ref{conclusion} summarizes our findings and discusses directions for future research.

\section{Preliminary}\label{Preliminary}

\subsection{Fisher Information Matrix and Cramér-Rao Inequality}\label{sec:fi-crlb}

In this section, we introduce the key theoretical tools for our analysis: the Fisher information matrix (FIM) and the Cramér-Rao lower bound (CRLB).
Let the parameters we wish to estimate be a multi-dimensional vector $\boldsymbol{\theta}=(\theta_0,\dots,\theta_{k-1})\in\mathbb{R}^k$, and let  $X$ be a random variable with a probability density function $f(x\mid\boldsymbol{\theta})$. The dependence of $f(x\mid\boldsymbol{\theta})$ on $\boldsymbol{\theta}$ is the key to determining the estimation precision. 

FIM is a metric that quantifies the sensitivity of the probability distribution to changes in $\boldsymbol{\theta}$. FIM is defined as:
\[
\mathcal{I}_{i,j}(\boldsymbol{\theta})
  = \mathbb{E}_{X}\!\left[
     \frac{\partial}{\partial \theta_i} \ln f(X|\boldsymbol{\theta}) \frac{\partial}{\partial \theta_j} \ln f(X|\boldsymbol{\theta})
    \right]
  \in\mathbb{R}^{k\times k}.
\]
The larger the components of the FIM, the more sensitive the probability distribution is to changes in its parameters, and we can expect a more precise estimation.
For independent random variables $X$ and $Y$, the FIM is additive: $\mathcal{I}_{X,Y}(\boldsymbol{\theta}) = \mathcal{I}_{X}(\theta) + \mathcal{I}_{Y}(\boldsymbol{\theta})$.

The FIM allows us to quantify the limit on the estimation precision for any unbiased estimator $\hat{\boldsymbol{\theta}}$. This is known as the Cramér-Rao Lower Bound (CRLB) and is expressed using the covariance matrix of $\hat{\boldsymbol{\theta}}$, $\operatorname{Cov}(\hat{\boldsymbol{\theta}})$, as
\[
\operatorname{Cov}(\hat{\boldsymbol{\theta}})
   \ \succeq\ \mathcal{I}(\boldsymbol{\theta})^{-1}.
\]
Here, the matrix inequality $A \succeq B$ means that $A-B$ is positive semidefinite (PSD). A well-known property that follows from the CRLB is the following inequality:
\begin{equation}
    \mathbb{E}[(\theta_i-\hat{\theta_i})^2] \ge \left(\mathcal{I}(\boldsymbol{\theta})^{-1}\right)_{i,i}.\label{eq:crlb}
\end{equation}

\subsection{Problem Setup}
Let $U$ be a unitary operator with the spectral decomposition $U=\sum^{L-1}_{l=0} e^{i\theta_l} \ket{\psi_l} \bra{\psi_l}$, where $\ket{\psi_l}$ is an eigenstate of $U$ and $\theta_l$ is the corresponding eigenphase. We consider the task of estimating a specific phase $\theta_i$ from a given input state $\ket{\psi}$ and the operator $U$. We assume that the input state $\ket{\psi}$ can be expressed using the eigenstates of $U$ as:
\begin{equation}
    \ket{\psi} = \sum^{L-1}_{l=0} \beta_{l}\ket{\psi_l}, \notag
\end{equation}
 where  $\beta_l$ is a complex number. For convenience, we define the quantity $c_l := \left| \braket{\psi}{\psi_l} \right|^{2}$ represents the overlap between the input state and  $l$-th eigenstate. These coefficients are central to determining the precision of the estimation.

To estimate phases of $U$, we assume the ability to implement the controlled-$U^t$ for an integer $t$. QPE consists of the following three steps: 1. Determine the sequence of integers $\{t_k \}^{N_{\text{t}}}_{k=1}$ to be used as the exponents in the controlled-$U^{t_k}$ gates.  2. For each $\{t_n \}$, execute the corresponding quantum circuit and perform a measurement $N_{\text{s}}$ times. 3. Estimate the phases using classical post-processing on the $N = N_{\text{s}}N_{\text{t}}$ measurement outcomes.

Note that QPE protocols are generally classified into two categories, adaptive and non-adaptive schemes, depending on the strategy used in Step 1.
In adaptive schemes, the $t_k$ is chosen sequentially based on the measurement outcomes obtained in previous steps.
In contrast, in non-adaptive schemes, the $t_k$ is determined independently of the measurement results.
In this work, we focus on the latter, non-adaptive protocols.

There are two key metrics to quantify the computational costs: the maximum number of applications of controlled-$U$, $T=\max_{k} |t_k|$, and the total number of applications of controlled-$U$, $t_{\text{total}}=N_{\text{s}}\sum^{N_{\text{t}}}_{k=1} |t_k|$. Note that $T$ and $t_{\text{total}}$ approximately represent the circuit depth and the total runtime of the algorithm, respectively.

\subsection{QFT-QPE}\label{preliminary_qft}
\begin{figure}[h]
  \centering
  \graphicspath{{./Figure/}}
  \includegraphics[width=0.8\hsize]{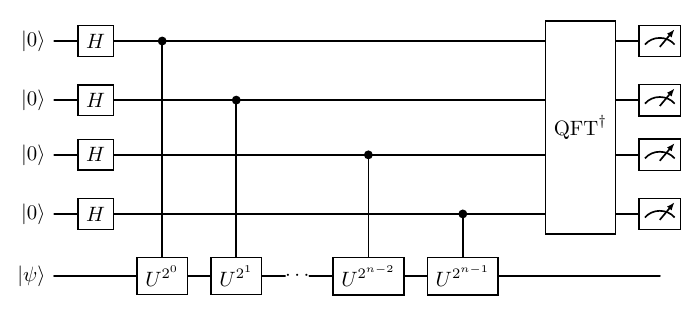}
  \caption{Quantum circuit for QFT-QPE. The upper $n$ qubits are referred to as ancillary qubits.}
  \label{fig:QFTQPE_kairo}
\end{figure}

In this section, we briefly review the QFT-QPE algorithm, outlining its quantum circuit and the resulting measurement probability distribution. Textbook QFT-QPE uses the circuit shown in Fig.~\ref{fig:QFTQPE_kairo}.
To achieve an estimation precision of $\mathcal{O}(1/2^n)$, we use a circuit with an n-qubit ancillary qubits as shown in Fig.~\ref{fig:QFTQPE_kairo}. For this configuration, the cost $T$ is given by
\[
T \;=\; 2^{0} + \cdots + 2^{n-1} \;=\; 2^{n} - 1.
\]
When we measure the ancillary qubits at the end of the circuit, we obtain an n-bit string. Interpreting this string as an integer $y\in\{0,\dots,2^n-1\}$, the probability of obtaining $y$ is given by:
\begin{align}
p(y \mid \boldsymbol{\theta}, \boldsymbol{c})
&= 
 \sum^{L-1}_{l=0} \frac{c_l}{2^{2n}}\, D^2_{2^n}\!\bigl(\phi_{l,y}\bigr),
\label{probqft}
\end{align}
where $\phi_{i,y}=\theta_i-\tfrac{2\pi y}{2^n}$, and $D_{2^n}(x)$ is the $2^n$-th order Dirichlet kernel, defined as $D_{2^n}(x)=\sin(2^nx/2)/\sin(x/2)$~\cite{dirichlet_qft}.

Although tapering window functions can mitigate the spectral leakage of the Dirichlet kernel~\cite{Kaiser, tapering2024optimal}, they inherently broaden the main lobe of the distribution, leading to a reduction in Fisher information~\cite{qftqpe_fisher}. Since our primary goal is to derive the fundamental performance limits, we focus exclusively on the unmodified Dirichlet kernel case in this study.

We sample $y$ from the distribution given in Eq.~\eqref{probqft} and perform classical post-processing to estimate $\theta_i$.
In many existing studies, this process involves executing the circuit shown in Fig.~\ref{fig:QFTQPE_kairo} multiple times to sample from the same output distribution~\cite{2409.15752}.
Therefore, in this work, we set $N_{\text{t}}=1$.
Consequently, the cost $t_{\text{total}}$ for QFT-QPE is given by
\begin{equation}
    t_{\text{total}} = N_{\text{s}} T.
\end{equation}

\subsection{HT-QPE}

\begin{figure}[h]
  \centering
  \graphicspath{{./Figure/}}
  \includegraphics[width=0.8\hsize]{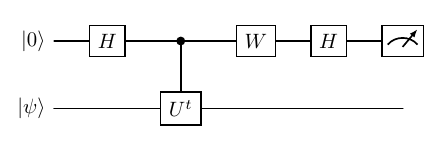}
  \caption{Quantum circuit for HT-QPE.}
  \label{fig:HTQPE_kairo}
\end{figure}

In this section, we briefly review HT-QPE, outlining its quantum circuit and the resulting measurement probability distributions. Figure~\ref{fig:HTQPE_kairo} shows the quantum circuit for HT-QPE. When $W=I$, we can obtain information about the real part of $\bra{\psi} U \ket{\psi }$, and when $W=S^{\dagger}$, we can obtain information about the imaginary part.
Given an input state $\ket{\psi}$, a unitary matrix $U$, and an integer $t$. The probability of obtaining the outcome $h \in \{0, 1\}$ from the measurement of the ancillary qubit for the real and imaginary parts, $p_{\text{Re}}(h|\boldsymbol{\theta},\boldsymbol{c})$ and $p_{\text{Im}}(h|\boldsymbol{\theta},\boldsymbol{c})$, as follows:
\begin{equation}
    p_{\text{Re}}(0|\boldsymbol{\theta},\boldsymbol{c}) = \frac{1+\Re(\bra{\psi}U^{t}\ket{\psi})}{2} = \frac{1+\sum^{L-1}_{l=0}c_l\cos(t \theta_l)}{2},\label{ht_re_prob_0}
\end{equation}
\begin{equation}
    p_{\text{Re}}(1|\boldsymbol{\theta},\boldsymbol{c}) = \frac{1-\Re(\bra{\psi}U^{t}\ket{\psi})}{2} = \frac{1-\sum^{L-1}_{l=0}c_l\cos(t\theta_l)}{2}.\label{ht_re_prob_1}
\end{equation}

\begin{equation}
    p_{\text{Im}}(0|\boldsymbol{\theta},\boldsymbol{c}) = \frac{1+\Im(\bra{\psi}U^{t}\ket{\psi})}{2} = \frac{1+\sum^{L-1}_{l=0}c_l\sin(t \theta_l)}{2},\label{ht_im_prob_0}
\end{equation}
\begin{equation}
    p_{\text{Im}}(1|\boldsymbol{\theta},\boldsymbol{c}) = \frac{1-\Im(\bra{\psi}U^{t}\ket{\psi})}{2} = \frac{1-\sum^{L-1}_{l}c_l\sin(t\theta_l)}{2}.\label{ht_im_prob_1}
\end{equation}

When estimating phases from the probability distributions in Eqs.~\eqref{ht_re_prob_0}-\eqref{ht_im_prob_1}, using only a single integer $t$ (i.e., $N_{\text{t}} = 1$) is generally impossible due to aliasing issues. Therefore, HT-QPE protocols typically use a sequence of different $\{t_k \}^{N_{\text{t}}}_{k=1}$ (i.e., $N_{\text{t}}>1$).
The sequence $\{t_k \}^{N_{\text{t}}}_{k=1}$ can follow two protocols for determining the sequence $\{t_k \}^{N_{\text{t}}}_{k=1}$: a stochastic protocol, in which the values are chosen probabilistically, and a deterministic protocol, in which they are fixed.

Representative examples of stochastic protocols include quantum multiple eigenvalue Gaussian filtered search (QMEGS)~\cite{2402.01013} and quantum phase estimation by compressed sensing (CSQPE)~\cite{CS}.
In QMEGS~\cite{2402.01013}, we determine $\{ t_k\}^{N_{\text{t}}}_{k=1}$ by sampling from a truncated normal distribution with $|t|\le T$:
\begin{equation}
  \frac{1}{C\sqrt{2\pi T^2}}\exp(-\frac{t^2}{2T^2})\textbf{1}_{[- T,T]}(t),
\end{equation}
where $C$ is a normalization factor.
In CSQPE~\cite{CS}, we determine $\{ t_k\}^{N_{\text{t}}}_{k=1}$ by sampling from a uniform distribution over the integers $t\in \left[1,T \right]$.

A representative example of a deterministic protocol is quantum complex exponential least squares (QCELS)~\cite{QCELS}.
In QCELS, we determine $\{ t_k\}^{N_{\text{t}}}_{k=1}$ as an arithmetic progression with a step size of $\Delta t = T/N_t$:
\begin{equation}
    t_k = k\Delta t , \quad (k = 1, \dots , N_t).
\end{equation}

We define the total cost $t_{\text{total}}$ for HT-QPE for both stochastic and deterministic protocols as follows:
\begin{equation}
    t_{\text{total}} = 2N_{\text{s}} \mathbb{E}\left[\sum^{N_{\text{t}}}_{k=1}|t_k|\right] ,\label{def:t_total}
\end{equation}
where the factor of 2 arises from measuring both the real and imaginary components.
Note that this single definition encompasses both stochastic and deterministic protocols; a deterministic protocol can be viewed as sampling from a probability distribution that is exclusively concentrated on the specific, predetermined values of $\{ t_k \}^{N_t}_{k=1}$.

As shown in the Appendix~\ref{app:gamma_chi}, $t_{\text{total}}$ for these protocols can be expressed in the form of $t_{\text{total}}=\gamma NT$, where $\gamma$ is a constant that depends on the specific protocol.
The values of $\gamma$ for each algorithm are shown in Table~\ref{tab:gamma}.

\begin{table}[h]
  \caption{The values of $\gamma$ for QMEGS, CSQPE, and QCELS.}
  \label{tab:gamma}
  \centering
  \renewcommand{\arraystretch}{1.8}
  \begin{tabular}{ccc}
    \toprule
    \textbf{QMEGS} &
    \textbf{CSQPE} &
    \textbf{QCELS} \\
    \midrule
    $\displaystyle
       \frac{2}{C}\!\left[1-\frac{1}{\sqrt{e}}\right]
       \sqrt{\frac{2}{\pi}} \approx 0.92$
      &
    $1$
      &
    $1$ \\[2ex]
    \bottomrule
  \end{tabular}
\end{table}

\section{Derivation of the lower bound on the cost product using FIM and CRLB}\label{Theory}

In this section, we formulate the theoretical limit of the estimation precision in QFT-QPE and HT-QPE using FIM and the CRLB. The analysis proceeds as follows: we first derive the general form of the FIM and the cost lower bound common to QFT-QPE and HT-QPE. Based on this unified framework, we then explicitly calculate the FIM and evaluate the cost scaling for QFT-QPE and HT-QPE, respectively.

First, we define the parameters for the FIM in our problem setting. We treat not only the phases we wish to estimate, $\boldsymbol{\theta}$, but also the overlaps between the input state and the eigenstates, $\textbf{c}$, as unknown parameters. We do this because in many applications, $\boldsymbol{c}$ is often not known in advance, and its uncertainty affects the estimation precision through the probability distribution.
The FIM with respect to these parameters, denoted as $\mathcal{I}(\boldsymbol{\theta}, \textbf{c})$, naturally adopts the following block matrix structure:
\begin{equation}
    \mathcal{I}(\boldsymbol{\theta}, \textbf{c}) = 
    \begin{pmatrix} 
        \mathcal{I}^{\theta\theta} & \mathcal{I}^{\theta c} \\ 
        \mathcal{I}^{c\theta} & \mathcal{I}^{cc} 
    \end{pmatrix},
    \label{eq:fim_block_structure}
\end{equation}
where the components of each block matrix are given for indices $i,j\in\{0,\dots,L-1 \}$. Note that $\mathcal{I}(\boldsymbol{\theta,c}|t_k)$ for HT-QPE accounts for the measurements of both the real and imaginary parts.

For QFT-QPE and HT-QPE, since the sampling at each $t_k$ is independent, we can obtain the total FIM by summing the FIMs from each time step. Similarly to the definition in~\eqref{def:t_total}, we define the total FIM $\mathcal{I}_{\text{total}} (\boldsymbol{\theta,c})$ as follows:
\begin{align}
    \mathcal{I}_{{\text{total}}}(\boldsymbol{\theta,c})
    &=
    \mathbb{E}\left[N_{\text{s}}\sum^{N_{\text{t}}}_{k=1} \mathcal{I}(\boldsymbol{\theta,c}|t_k)\right].\label{eq:fim_total}
\end{align}

Using the total FIM derived above, the CRLB for the MSE of any unbiased estimator $\hat{\theta}_i$ is given by the Eq.~\eqref{eq:crlb}, provided that $\mathcal{I}_{\text{total}}$ is non-singular.
Note that this general bound is valid only when the FIM is invertible. A typical case where the FIM becomes singular is when the eigenphases are degenerate. In such a case, the parameters become unidentifiable, and the problem shifts to estimating the subspace or the sum of probabilities, which corresponds to a fundamentally different estimation task. Therefore, we exclude singular cases from the scope of this work.

Even provided that the FIM is non-singular, deriving an explicit closed-form expression for the full inverse matrix $\mathcal{I}_{\text{total}}^{-1}$ is analytically complicated, which obscures clear physical insights.
Instead, we utilize the inequality introduced in Eq. (1): $(\mathcal{I}_{\text{total}}^{-1})_{i,i} \ge 1/(\mathcal{I}_{\text{total}})_{i,i}$.
While this bound is looser than the exact CRLB in general, we proceed under the assumption that it provides a sufficiently accurate approximation in the asymptotic regime $T \gg \Delta_{\min}^{-1}$. We numerically demonstrate the validity of this approximation in Sec.~\ref{main_results}.

Under this diagonal approximation, as we will discuss later, we observe a scaling structure in the Fisher information for non-adaptive QPE protocols.
Specifically, in the asymptotic regime, the diagonal elements of the total Fisher information for both methods scale proportionally to $NT^2$. The proportionality constant, denoted by $g_i$, is defined as:
\begin{equation}
    g_i = \frac{(\mathcal{I}_{\text{total}})_{i,i}}{NT^2}.\label{g_i}
\end{equation}
Hereafter, we refer to this as the information acquisition efficiency.
As shown in Appendix~\ref{app:qftqpe_gi_scaling},~\ref{app:htqpe_gi_scaling}, by analyzing $g_i$ for both QFT-QPE and HT-QPE, we reveal that their scalings with respect to $c_i$ are
\begin{equation}
    g^{\text{QFT}}_{i} =  \Theta(c_i),\label{g_i_qft}
\end{equation}

\begin{equation}
    g^{\text{HT}}_{i} = \Theta(c_i^2).\label{g_i_ht}
\end{equation}
This difference in $g_i$ dictates the distinct performance characteristics of the two algorithms.

By substituting this expression into the CRLB and using the runtime relation $t_{\text{total}} = \gamma N T$, we derive a lower bound for the cost product:
\begin{equation}
    T t_{\text{total}} \ge \frac{\gamma}{g_i \, \mathbb{E}[(\theta_i-\hat{\theta_i})^2]}.\label{lower_bound}
\end{equation}
This inequality shows that the theoretical limit of the cost product is determined by the ratio $\gamma/g_i$.

Regarding Eq.~\eqref{lower_bound}, by comparing QFT-QPE and HT-QPE, we reveal a difference: the cost lower bound for HT-QPE scales inversely with the square of the overlap ($c_i^{-2}$), whereas that for QFT-QPE scales with $c_i^{-1}$.
This suggests that QFT-QPE is asymptotically more efficient in the regime of small initial overlap.

The scaling of the total runtime in quantum phase estimation (QPE) under constrained circuit depths is a subject of study in the development of quantum algorithms. Previous studies, for example, have identified the establishment of lower bounds on the total runtime under restricted circuit depths as a challenge for future research~\cite{Wang_QPE}. This study addresses this challenge by formulating cost lower bounds for specific circuit designs. Specifically, Eq.~\eqref{lower_bound} determines the lower bound of the total runtime $t_{\text{total}}$ when the circuit depth $T$ is constrained within the quantum circuit configurations of QFT-QPE and HT-QPE. Building upon these bounds, a key future direction will be to derive a lower bound that is independent of specific circuit structures, potentially by utilizing metrics such as the quantum Fisher information.

Next, we comment on the applicability of our framework to robust phase estimation (RPE)~\cite{RPE}. 
As discussed in Appendix~\ref{app:RPE}, RPE does not fit the present setting in which the total runtime admits a simple relation of the form $t_{\text{total}}=\gamma N T$ and the cost product lower bound can be expressed solely through the ratio $\gamma/g_i$ in Eq.~\eqref{lower_bound}. 
However, it is still possible to characterize the complexity of the cost product for RPE at the level of scaling, as summarized in Appendix~\ref{app:RPE}. 
We emphasize, however, that the Fisher-information-based diagonal approximation used in our analysis becomes substantially looser for RPE since RPE employs only a small number of $N_t$.

\section{Evaluation of CRLB and comparison of QPE methods}\label{main_results}

In this section, we numerically evaluate FIM and the cost lower bounds derived in ~\ref{Theory} to validate our theoretical findings and assess the efficiency of practical QPE algorithms.
Our analysis proceeds in three steps.
First, we verify the validity of the diagonal approximation of the FIM, which was assumed in our theoretical derivation, particularly in the asymptotic regime $T \gg \Delta_{\min}^{-1}$.
Second, we numerically confirm the scaling laws of the information acquisition efficiency $g_i$, specifically verifying its linear dependence on the overlap $c_i$ for QFT-QPE and quadratic dependence for HT-QPE.
Finally, we analyze the overlap dependence of the derived lower bounds for both QFT-QPE and HT-QPE to highlight their distinct scaling behaviors. We then compare these theoretical limits with the actual costs of representative algorithms to evaluate how closely current implementations approach the fundamental bounds.

\subsection{Numerical calculation settings}

In this section, we outline the numerical simulation settings used to validate our theoretical framework.
First, we define the distributions of the eigenphases and the overlaps, which are essential for the numerical evaluation of the FIM and the CRLB.
Second, to compare the derived theoretical lower bounds on the cost product $T t_{\text{total}}$ with actual performance, we specify the representative practical algorithms employed in our benchmarks.

\begin{figure}[h]
\centering
\graphicspath{{./Figure/}}
\includegraphics[width=0.8\hsize]{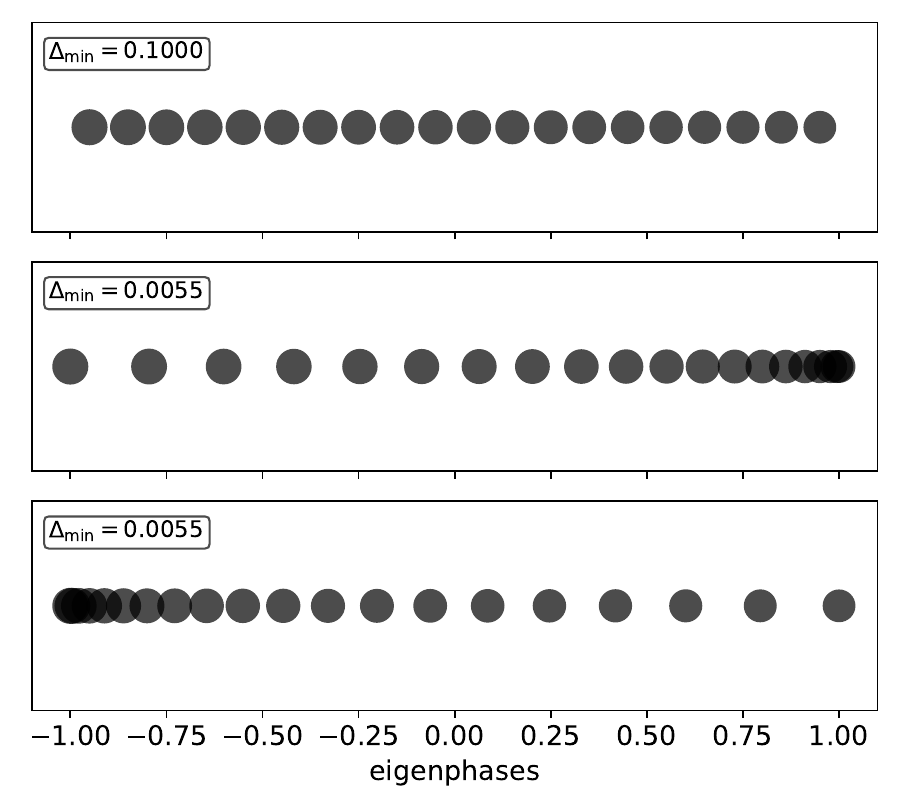}
\caption{Phase distributions for $L=20$.}
\label{spectrum_distribution_tau_u}
\end{figure}

We numerically evaluate the FIM for QFT-QPE and HT-QPE based on its analytical expressions (Eqs.~\eqref{eq:ht_FIM_thetatheta}-\eqref{eq:ht_FIM_ctheta}) rather than simulating quantum circuits.
To this end, we consider the following three cases for the distributions of $\boldsymbol{\theta}$: the uniform distribution (Eq.~\eqref{uniform distribution}), the head-dense distribution (Eq.~\eqref{head-dense distribution}) and the tail-dense distribution (Eq.~\eqref{tail-dense distribution}):
\begin{equation}
    \theta_i = -1 + \frac{2i+1}{L} ,\quad (i=0,\dots,L-1), \label{uniform distribution}
\end{equation}
\begin{equation}
    \theta_i = -1 + 2\left(\frac{i}{L-1}\right)^2 ,\quad (i=0,\dots,L-1), \label{head-dense distribution}
\end{equation}
\begin{equation}
    \theta_i = -1 + 2\left(\frac{L-1-i}{L-1}\right)^2 ,\quad (i=0,\dots,L-1). \label{tail-dense distribution}
\end{equation}
These distributions are visualized in Fig.\ref{spectrum_distribution_tau_u} for $L=20$.
As for $\boldsymbol{c}$, we assume a distribution similar to the one in~\cite{CS}:
\begin{equation}
  \boldsymbol{c} = \bigl[c_0, c_1, \ldots, c_{L-1}\bigr],\notag
\end{equation}
\begin{equation}\label{eq:population}
  c_l = \frac{(1-\alpha)\alpha^{l}}{1-\alpha^{L}},\qquad 0 < \alpha < 1,
\end{equation}
which is similar to a geometric distribution. In the distribution for $\boldsymbol{c}$ defined in Eq.~\eqref{eq:population}, $c_0$ is the largest overlap coefficient when $0 < \alpha < 1$. In many applications of quantum algorithms, the main objective is to estimate the phase of the eigenstate with the largest contribution, like this one.
 Therefore, in this analysis, we focus on evaluating the estimation error of the eigenphase $\theta_0$ corresponding to $i=0$ and examine the relationship between $g_0$ and $c_0$ in detail. Focusing on the $i=0$ case is sufficient to understand the qualitative behavior, as other indices show a similar trend.
By examining these settings for $\boldsymbol{\theta}$ and $\boldsymbol{c}$, we comprehensively investigate the behavior of the FIM.

To validate the derived theoretical lower bounds against practical performance, we perform numerical simulations using representative algorithms for both schemes. For the QFT-QPE, we employ the Curve-fitted QPE~\cite{2409.15752}, incorporating modifications to the original proposal as detailed in Appendix~\ref{app:curve_fitting_implementation}. For the HT-QPE framework, we consider three established algorithms: QMEGS~\cite{2402.01013}, CSQPE~\cite{CS}, and QCELS~\cite{QCELS}. The detailed simulation parameters for these algorithms, such as $T$, $N_{\text{s}}$ and $N_{\text{t}}$ are summarized in Table~\ref{tab:settings_practical_qpe}.

\subsection{Validity of diagonal approximation}
\label{validity of diagonal approximation}

\begin{figure*}[t]
  \centering
  \begin{adjustbox}{max width=\textwidth,max totalheight=\textheight}
    \begin{tabular}{ccc}
      \subfloat[Uniform distribution\label{fig:uniform_dist_qft}]{
        \begin{minipage}[b]{.32\linewidth}
          \centering
          \includegraphics[width=\linewidth]{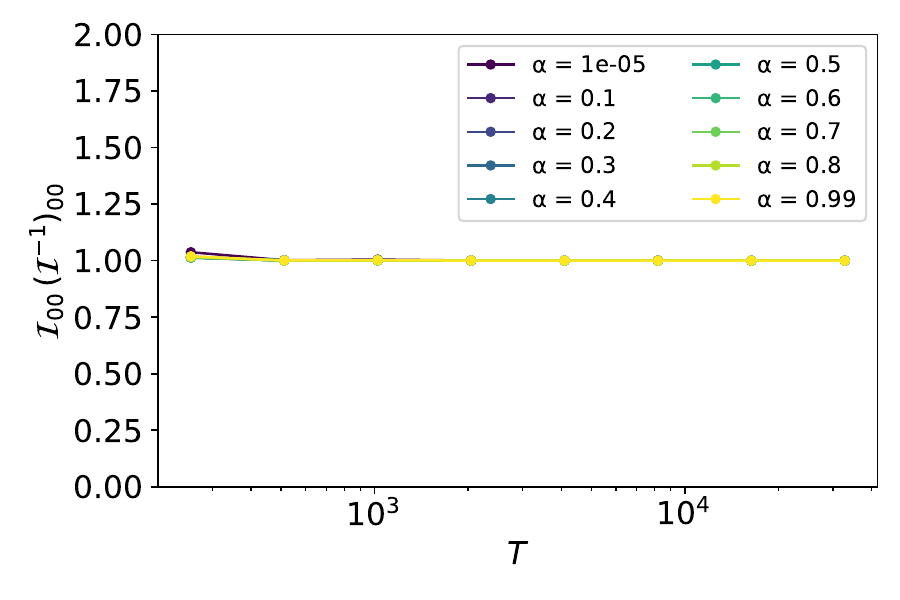}\\[-0.3em]
        \end{minipage}
      } &
      \subfloat[Tail-dense distribution\label{fig:tail_dense_qft}]{
        \begin{minipage}[b]{.32\linewidth}
          \centering
          \includegraphics[width=\linewidth]{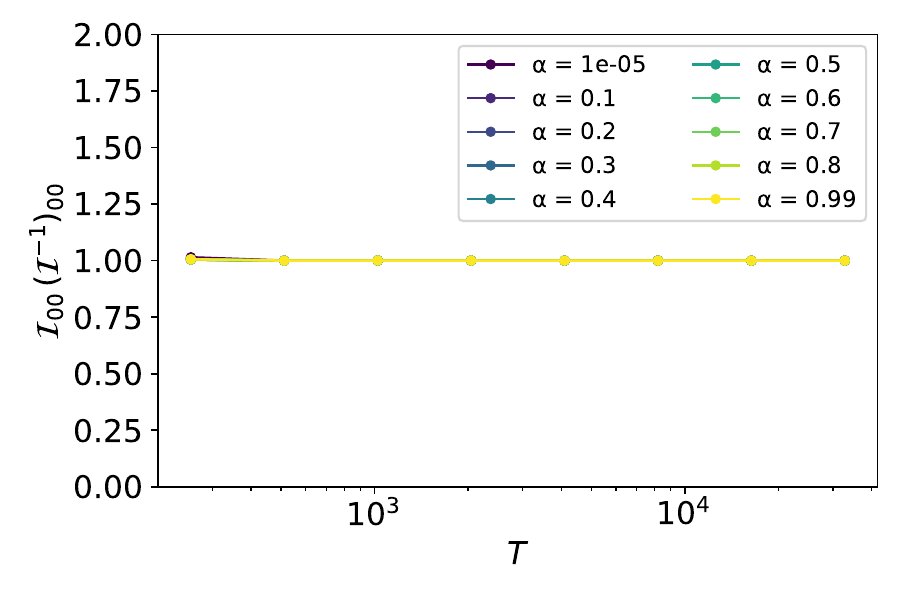}\\[-0.3em]

        \end{minipage}
      } &
      \subfloat[Head-dense distribution\label{fig:dense_head_qft}]{
        \begin{minipage}[b]{.32\linewidth}
          \centering
          \includegraphics[width=\linewidth]{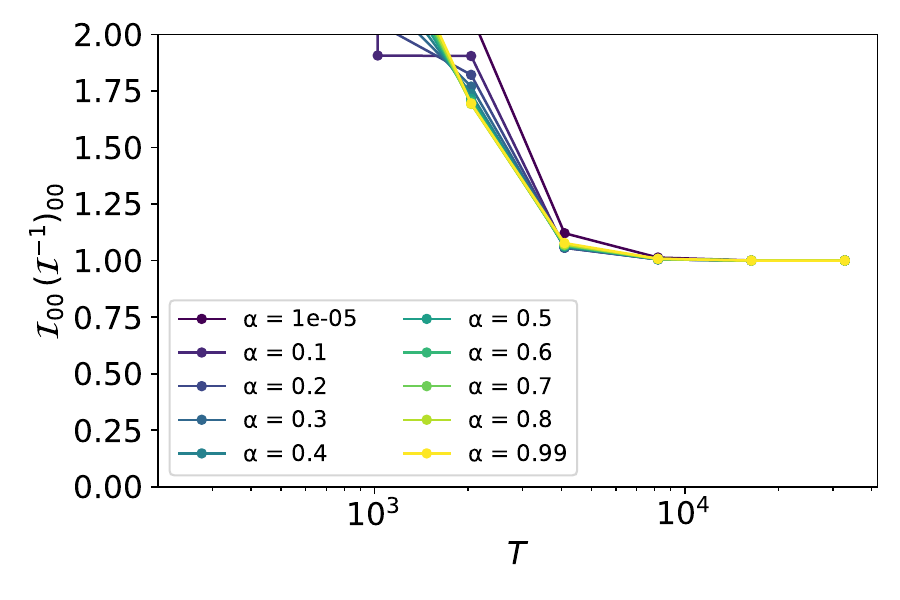}\\[-0.3em]
        \end{minipage}
      }
    \end{tabular}
  \end{adjustbox}
  \caption{Diagonal-FIM approximation check for QFT-QPE at $L=20$. (a) Uniform distribution, (b) Tail-dense distribution, and (c) Head-dense distribution.}
  \label{fig:diag_fim_grid_qft}
\end{figure*}

\begin{figure*}[t]
  \centering
  \begin{adjustbox}{max width=\textwidth,max totalheight=\textheight}
    \begin{tabular}{ccc}
      \subfloat[Uniform distribution\label{fig:uniform_dist}]{
        \begin{minipage}[b]{.32\linewidth}
          \centering
          \includegraphics[width=\linewidth]{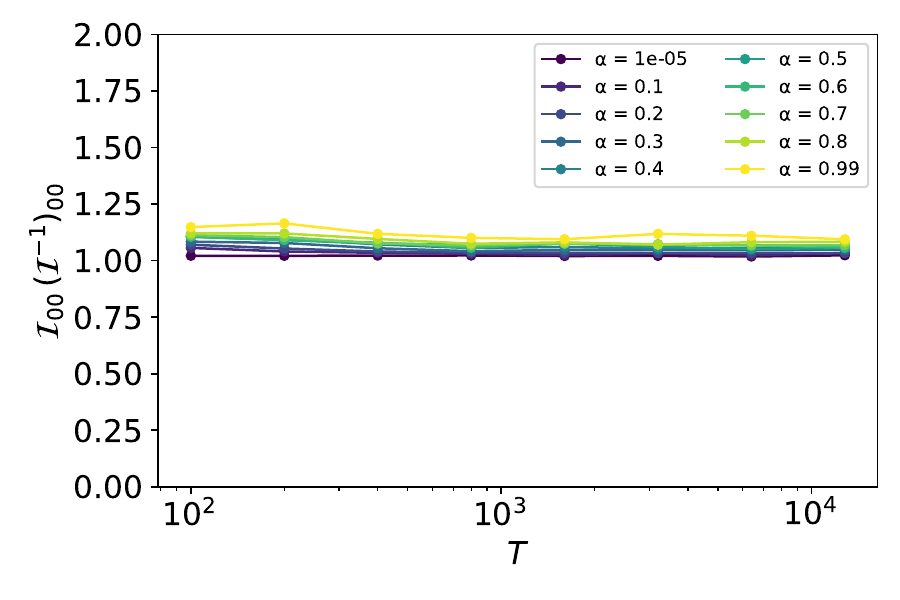}\\[-0.3em]
          \includegraphics[width=\linewidth]{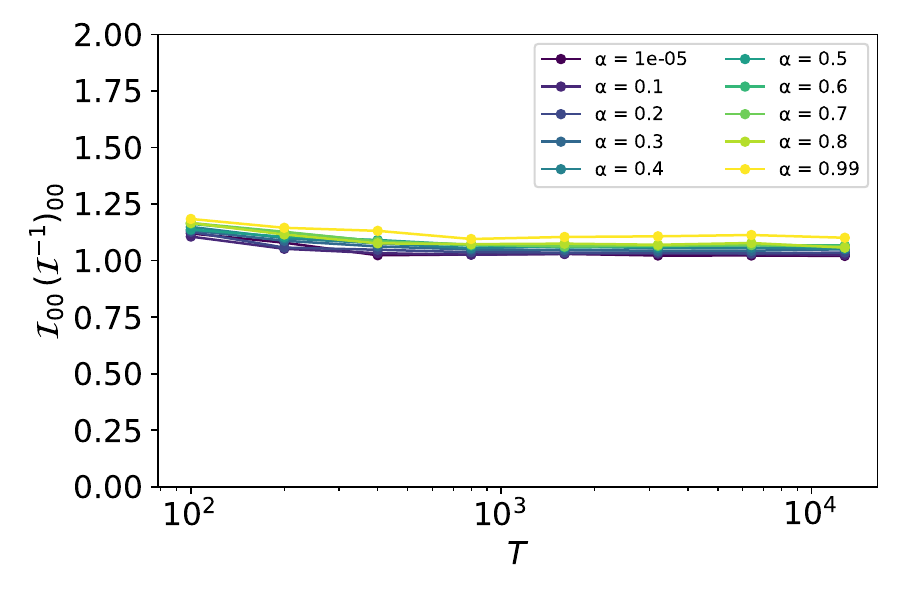}\\[-0.3em]
          \includegraphics[width=\linewidth]{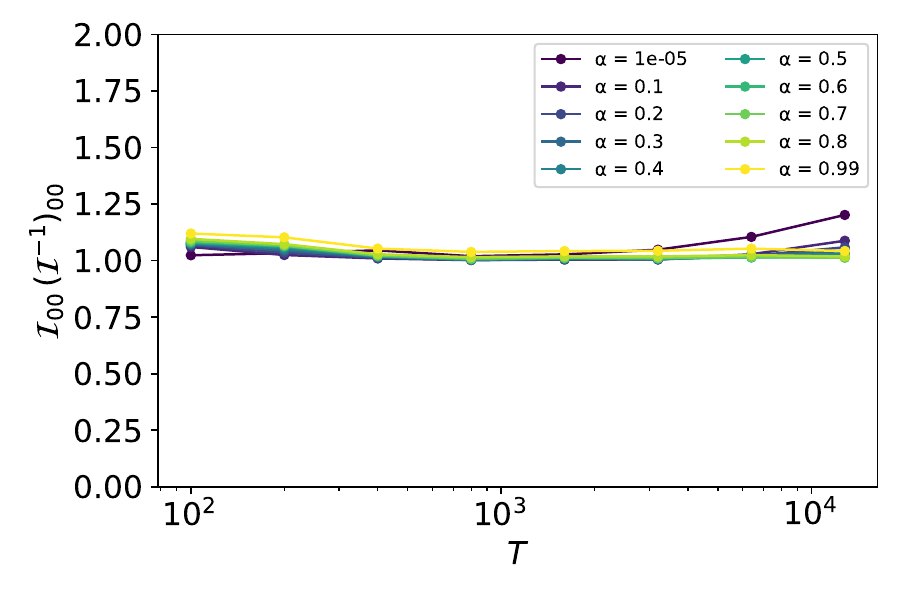}
        \end{minipage}
      } &
      \subfloat[Tail-dense distribution\label{fig:tail_dense}]{
        \begin{minipage}[b]{.32\linewidth}
          \centering
          \includegraphics[width=\linewidth]{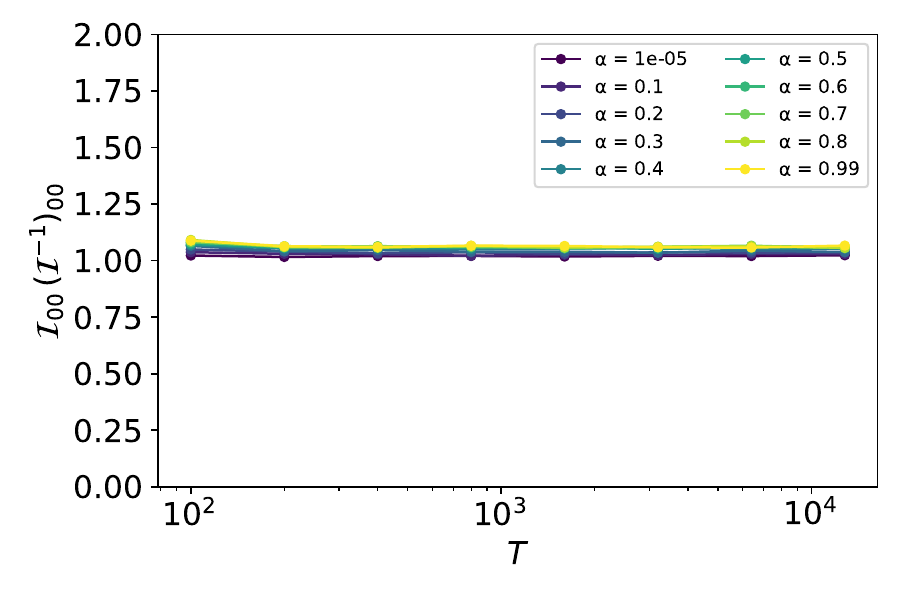}\\[-0.3em]
          \includegraphics[width=\linewidth]{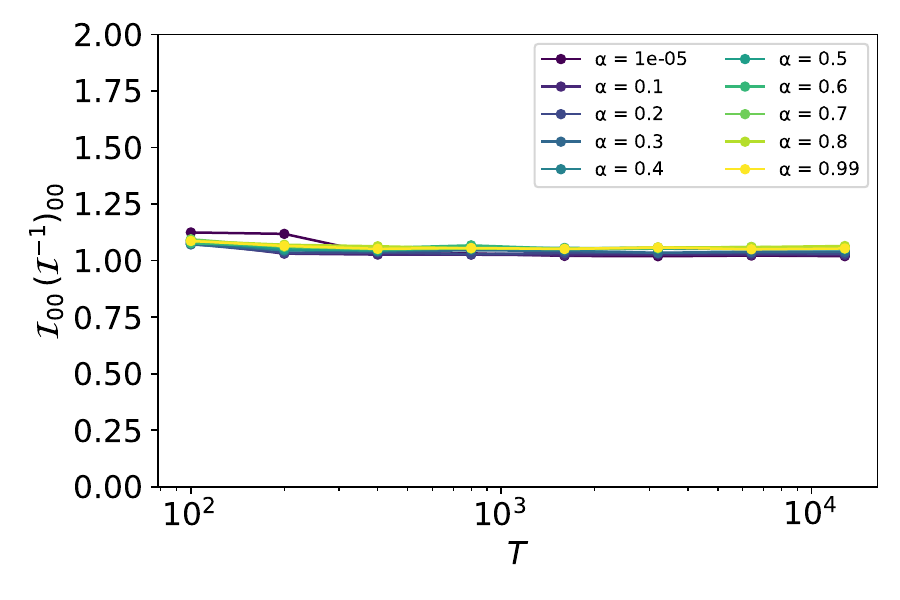}\\[-0.3em]
          \includegraphics[width=\linewidth]{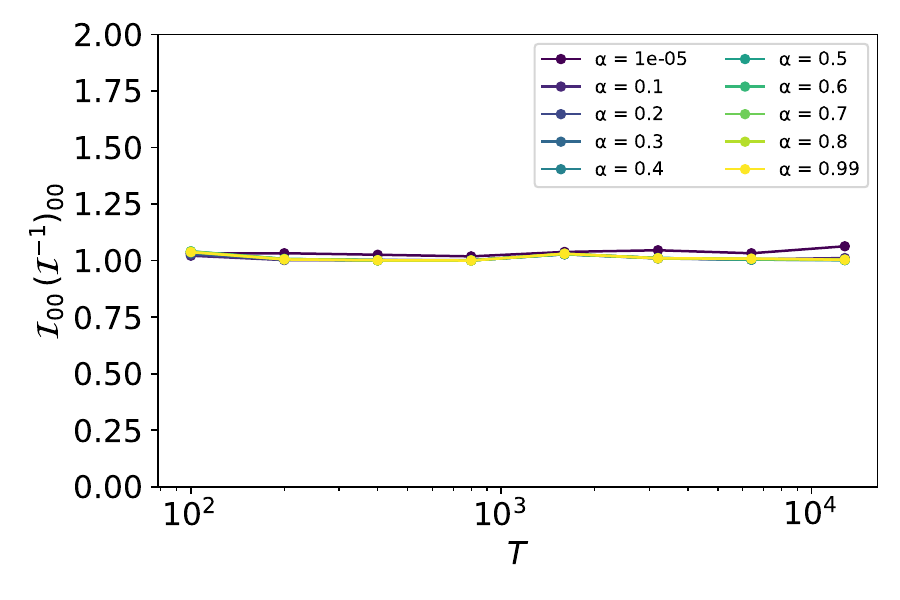}
        \end{minipage}
      } &
      \subfloat[Head-dense distribution\label{fig:head_dense}]{
        \begin{minipage}[b]{.32\linewidth}
          \centering
          \includegraphics[width=\linewidth]{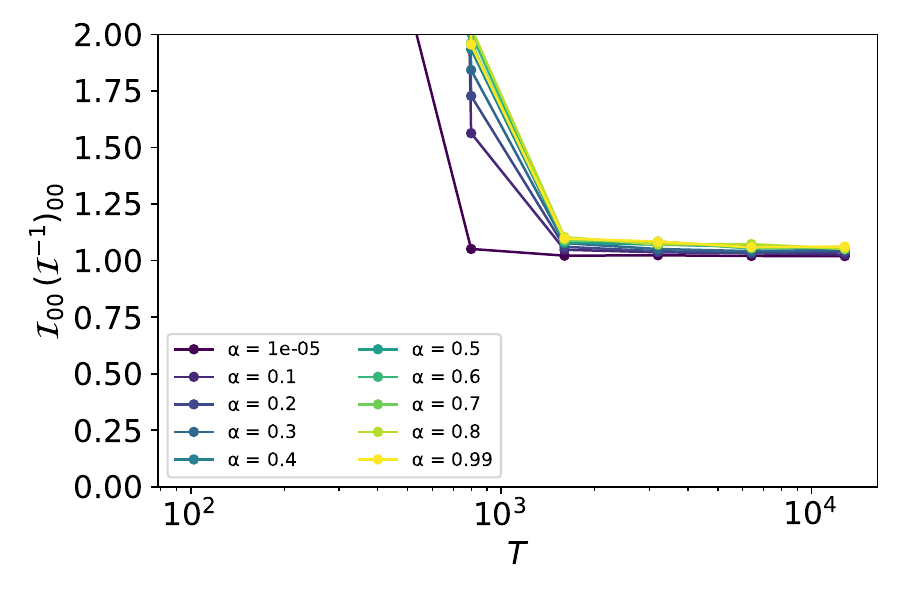}\\[-0.3em]
          \includegraphics[width=\linewidth]{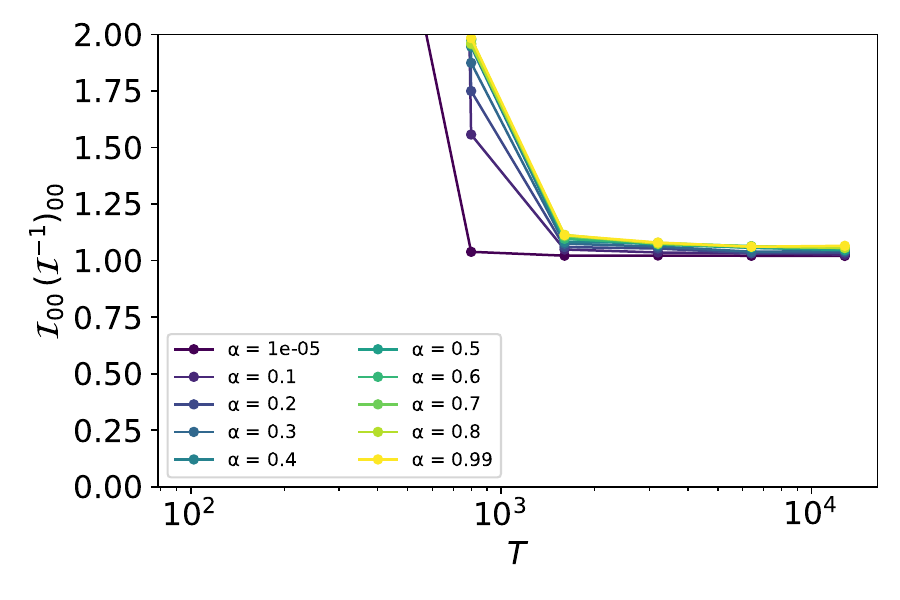}\\[-0.3em]
          \includegraphics[width=\linewidth]{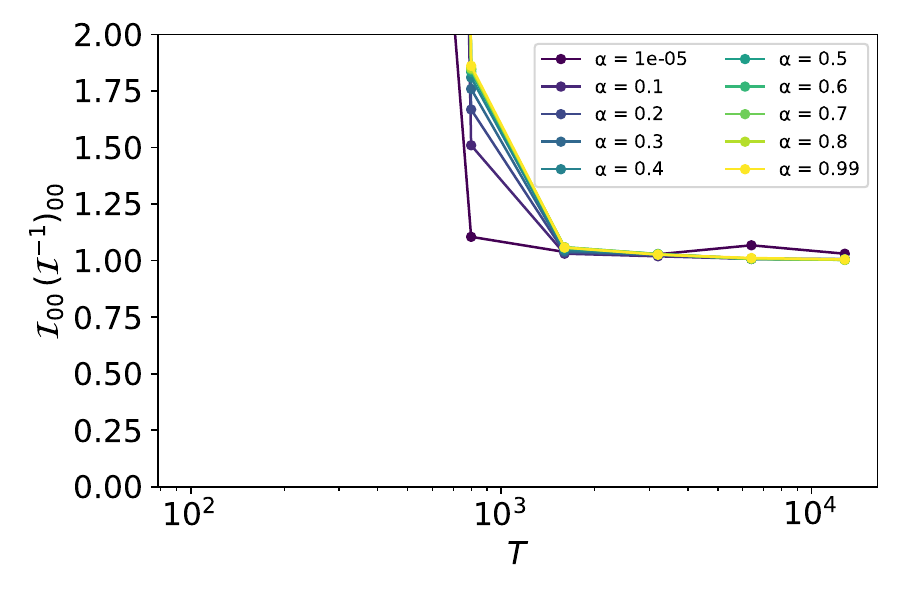}
        \end{minipage}
      }

    \end{tabular}
  \end{adjustbox}
  \caption{Diagonal-FIM approximation check at $L=20$. (a) Uniform distribution, (b) Tail-dense distribution, and (c) Head-dense distribution. In each subfigure the three stacked panels correspond to QMEGS (top), CSQPE (middle), and QCELS (bottom).}
  \label{fig:diag_fim_grid}
\end{figure*}

In this section, we numerically demonstrate that approximating the CRLB using only the diagonal elements of the FIM is valid under the conditions $\Delta_{\min}>0$ and $T \gg \Delta_{\min}^{-1}$.
Figures~\ref{fig:diag_fim_grid_qft} and \ref{fig:diag_fim_grid} show the $T$ dependence of the product $(\mathcal{I}_{\text{total}})_{0,0}\times (\mathcal{I}^{-1}_{\text{total}})_{0,0}$ for the uniform, tail-dense, and head-dense distributions, respectively. For sufficiently large $T$, the value of $(\mathcal{I}_{\text{total}})_{0,0}\times (\mathcal{I}^{-1}_{\text{total}})_{0,0}$ is indeed close to 1, confirming that the approximation using only the diagonal components for the evaluation of the CRLB is valid.

However, the plots also reveal that the approximation breaks down for smaller $T$, particularly in the head-dense distribution. This breakdown occurs precisely in the regime where the condition $T \gg \Delta_{\min}^{-1}$ is not satisfied. When $T$ is too small to resolve $\Delta_{\min}$, the information from different eigenphases becomes indistinguishable, leading to significant off-diagonal terms in the FIM. Since our primary interest lies in the large-$T$ regime required for high-precision estimation, we proceed with our cost analysis using the diagonal components of the FIM.

\subsection{Information acquisition efficiency}
\label{g_i_num}

In this section, we numerically demonstrate that the scaling laws derived in Sec.~\ref{Theory}, namely Eqs.~\eqref{g_i_qft} and \eqref{g_i_ht}, hold under the conditions $\Delta_{\min} > 0$ and $T \gg \Delta_{\min}^{-1}$. We numerically calculate $g_i$ based on Eq.~\eqref{g_i} using the analytical FIM expressions derived in Appendix~\ref{appendix_qft} and Appendix~\ref{appendix_ht}, focusing on the asymptotic regime $T \gg \Delta_{\text{min}}^{-1}$. The simulation settings are summarized in Table~\ref{tab:sim_param}.

\begin{figure*}[t]        
  \centering
  \begin{adjustbox}{max width=\textwidth}
    \subfloat[Uniform distribution\label{fig:htqpe_uniform}]{
      \begin{minipage}[b]{.32\linewidth}
        \centering
        \includegraphics[width=\linewidth]{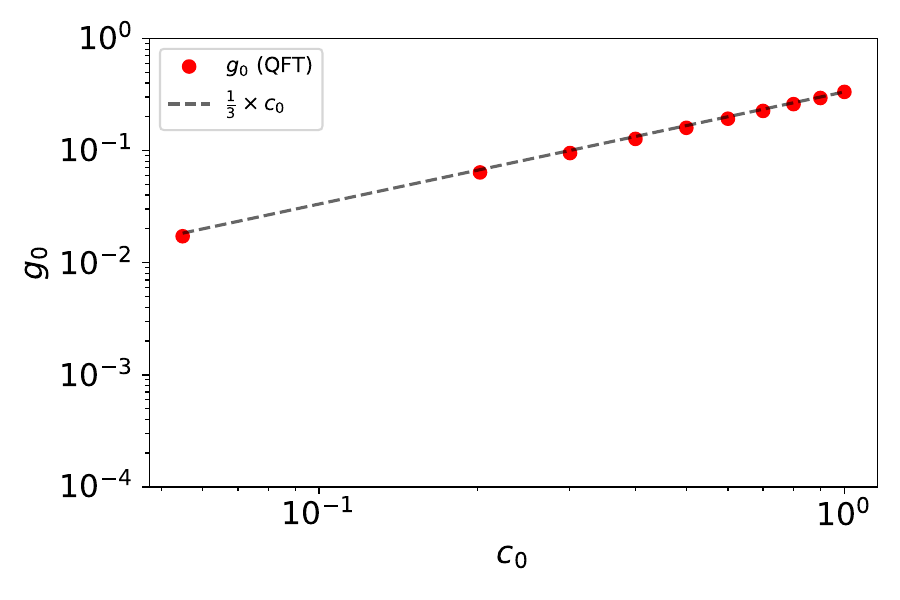}
      \end{minipage}
    }\hfill
    \subfloat[Head-dense distribution\label{fig:htqpe_head}]{
      \begin{minipage}[b]{.32\linewidth}
        \centering
        \includegraphics[width=\linewidth]{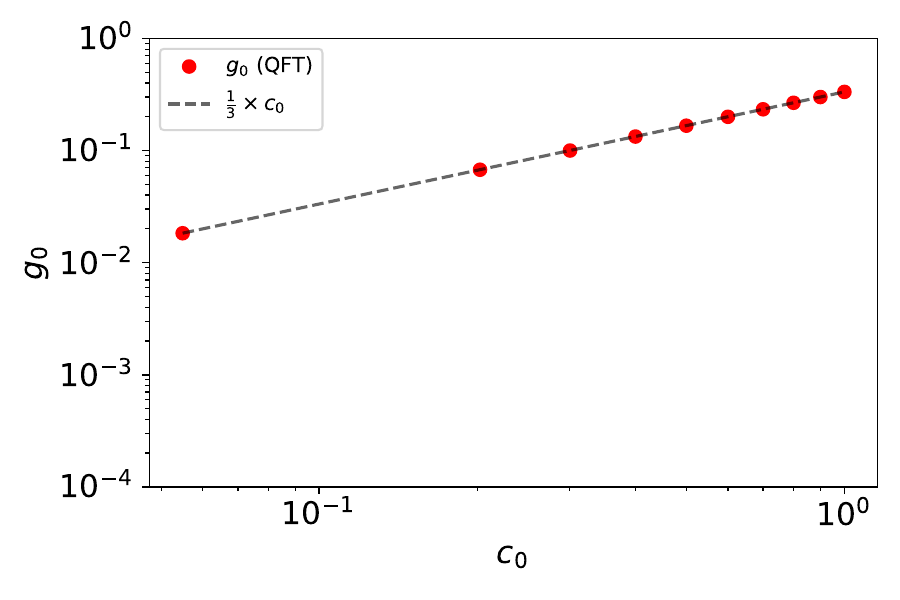}
      \end{minipage}
    }\hfill
    \subfloat[Tail-dense distribution\label{fig:htqpe_tail}]{
      \begin{minipage}[b]{.32\linewidth}
        \centering
        \includegraphics[width=\linewidth]{Figure/QFT-QPE_g_i_dense_head_20.pdf}
      \end{minipage}
    }
  \end{adjustbox}

  \caption{Dependence of $g_0$ on $c_0$ for QFT-QPE at $L=20$. (a) Uniform distribution, (b) Tail-dense distribution, and (c) Head-dense distribution.}
  \label{fig:qftqpe_g0_grid}
\end{figure*}

\begin{figure*}[t]        
  \centering
  \begin{adjustbox}{max width=\textwidth}
    \subfloat[Uniform distribution\label{fig:htqpe_uniform}]{
      \begin{minipage}[b]{.32\linewidth}
        \centering
        \includegraphics[width=\linewidth]{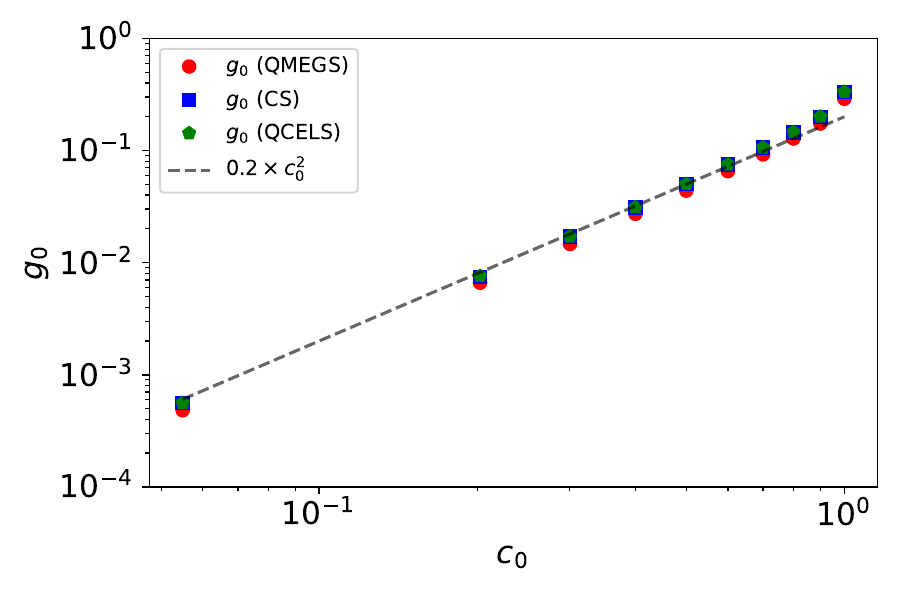}
      \end{minipage}
    }\hfill
    \subfloat[Head-dense distribution\label{fig:htqpe_head}]{
      \begin{minipage}[b]{.32\linewidth}
        \centering
        \includegraphics[width=\linewidth]{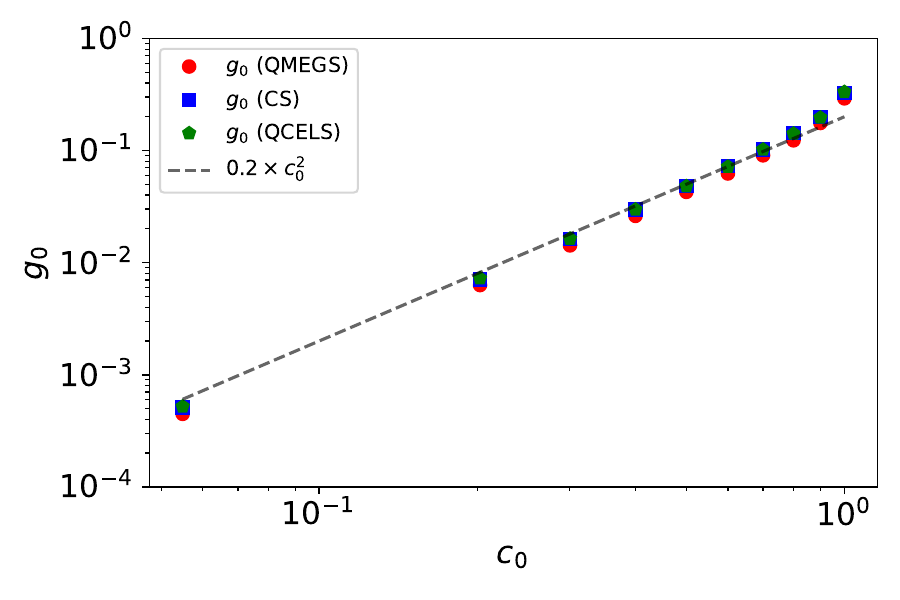}
      \end{minipage}
    }\hfill
    \subfloat[Tail-dense distribution\label{fig:htqpe_tail}]{
      \begin{minipage}[b]{.32\linewidth}
        \centering
        \includegraphics[width=\linewidth]{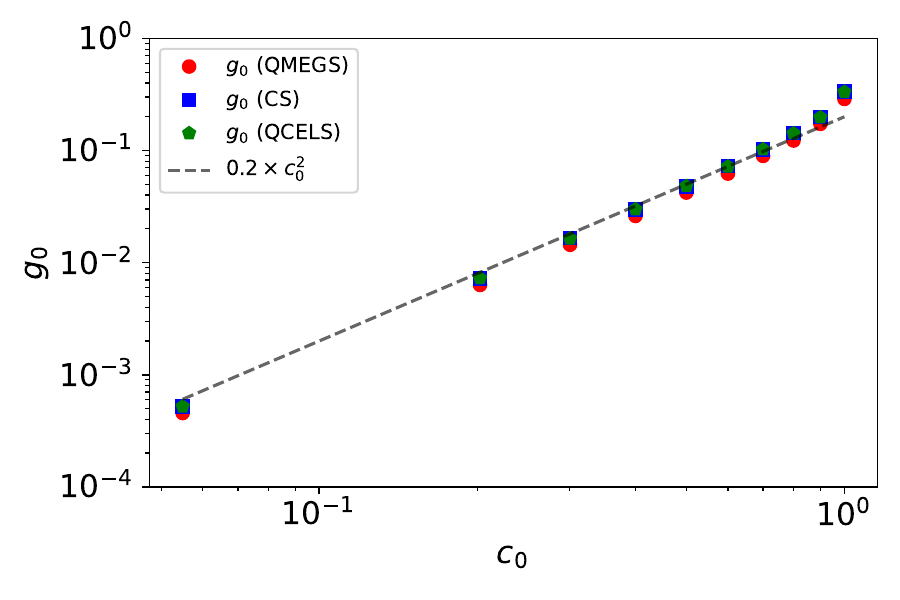}
      \end{minipage}
    }
  \end{adjustbox}

  \caption{Dependence of $g_0$ on $c_0$ for HT-QPE at $L=20$. (a) Uniform distribution, (b) Head-dense distribution, and (c) Tail-dense distribution.}
  \label{fig:htqpe_g0_grid}
\end{figure*}

Figures~\ref{fig:qftqpe_g0_grid} and \ref{fig:htqpe_g0_grid} show the dependence of $g_0$ on the overlap $c_0$ for the uniform distribution, head-dense distribution, and tail-dense distribution, respectively. For QFT-QPE, the numerical results in Fig.~\ref{fig:qftqpe_g0_grid} clearly exhibit a linear relationship between $g_0$ and $c_0$. This is consistent with the analytical prediction $g_0 \approx c_0/3$ derived in Eq.~\eqref{eq:g_i_qft}, confirming that the information efficiency scales linearly with the overlap.
Next, for HT-QPE, the results in Fig.~\ref{fig:htqpe_g0_grid} show a quadratic dependence on $c_0$. As seen in the graphs, for all considered protocols (QMEGS, CSQPE, and QCELS), the relation $g_0 = \mathcal{O}(c_0^2)$ holds. Note that while the scaling power is universally quadratic, the proportionality constant ($g_i \propto c^2_i$) depends on the specific method for determining the time sequence $\{t_k\}_{k=1}^{N_{\text{t}}}$.

These results numerically validate that the information acquisition efficiency of HT-QPE decays much faster than that of QFT-QPE as the overlap decreases. This fundamental difference supports the cost analysis discussed in the following section.

\begin{table}[h]
  \centering
  \footnotesize
  \setlength{\tabcolsep}{3pt} 
  \caption{Simulation settings: $T$, $N_{\text{t}}$, and $N_{\text{s}}$.}
  \label{tab:sim_param}
  \begin{tabular}{@{}l l c c@{}}
    \hline
    Algorithm & $T$ & $N_t$ & $N_s$ \\\hline
    QMEGS &
      $\{100,200,400,800,1600,3200,6400,12800\}$ &
      500 & 2 \\
    CSQPE &
      Same as above &
      500 & 2 \\
    QCELS &
      Same as above &
      100 & 200 \\
    QFT-QPE &
     $\{2^8-1,2^9-1,\dots,2^{15}-1\}$ &
     1 & 50 \\\hline
  \end{tabular}
\end{table}


\begin{figure*}[t]           
  \centering
  \begin{adjustbox}{max width=\textwidth}
    \subfloat[Uniform distribution\label{fig:gamma_uniform}]{
      \begin{minipage}[b]{.32\linewidth}
        \centering
        \includegraphics[width=\linewidth]{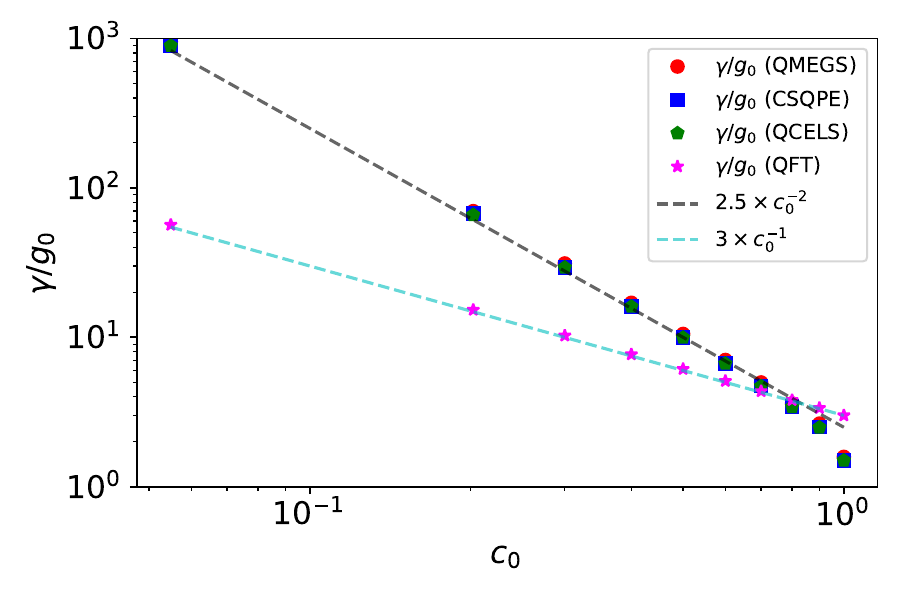}
      \end{minipage}
    }\hfill

    \subfloat[Tail-dense distribution\label{fig:gamma_tail}]{
      \begin{minipage}[b]{.32\linewidth}
        \centering
        \includegraphics[width=\linewidth]{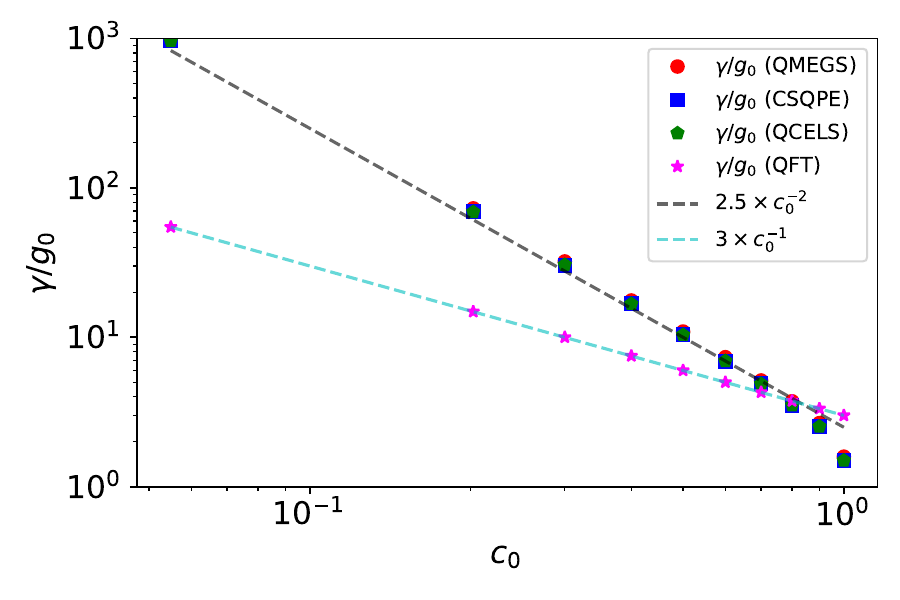}
      \end{minipage}
    }\hfill
    \subfloat[Head-dense distribution\label{fig:gamma_head}]{
      \begin{minipage}[b]{.32\linewidth}
        \centering
        \includegraphics[width=\linewidth]{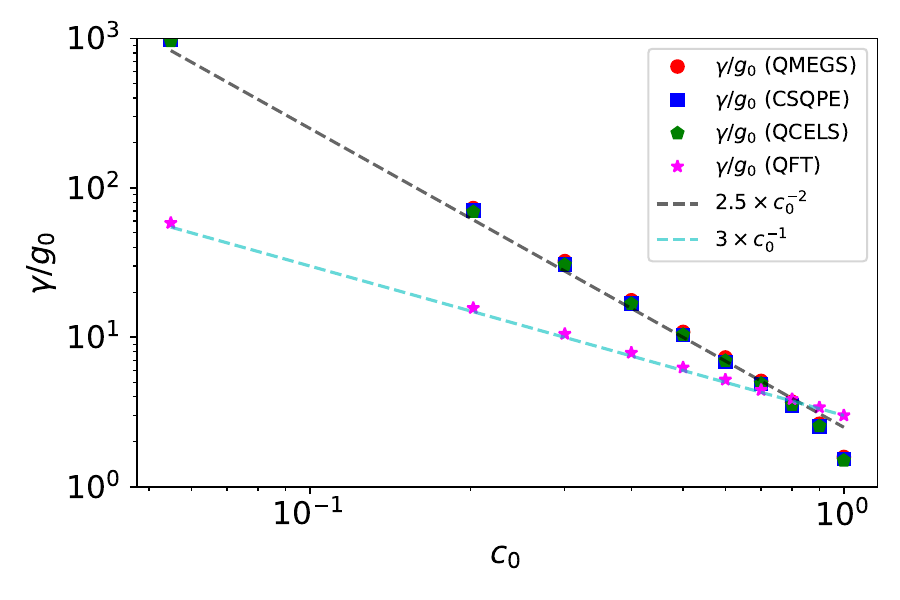}
      \end{minipage}
    }
  \end{adjustbox}

  \caption{Dependence of $\gamma/g_0$ on $c_0$ at $L=20$. (a) Uniform distribution, (b) Tail-dense distribution, and (c) Head-dense distribution.}
  \label{fig:gamma_grid}
\end{figure*}

\begin{figure*}[t]
  \centering
  \begin{adjustbox}{max width=\textwidth,max totalheight=\textheight}
    \begin{tabular}{ccc}
      \subfloat[Uniform distribution\label{fig:uniform_dist}]{
        \begin{minipage}[b]{.32\linewidth}
          \centering
          \includegraphics[width=\linewidth]{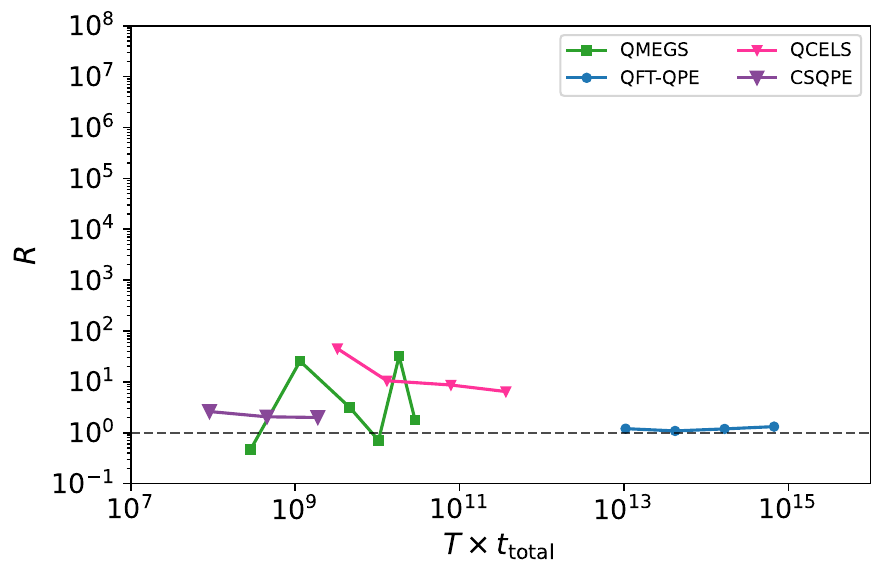}\\[-0.3em]
          \includegraphics[width=\linewidth]{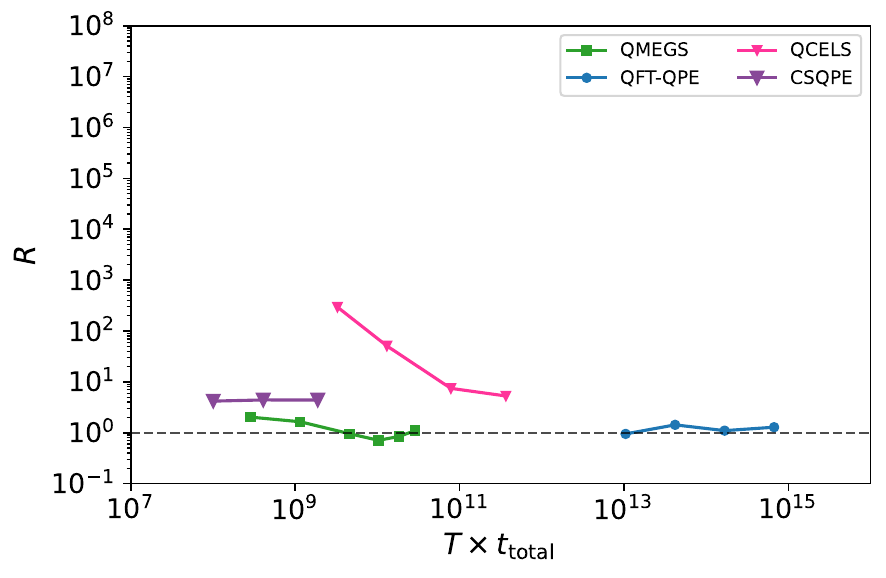}\\[-0.3em]
          \includegraphics[width=\linewidth]{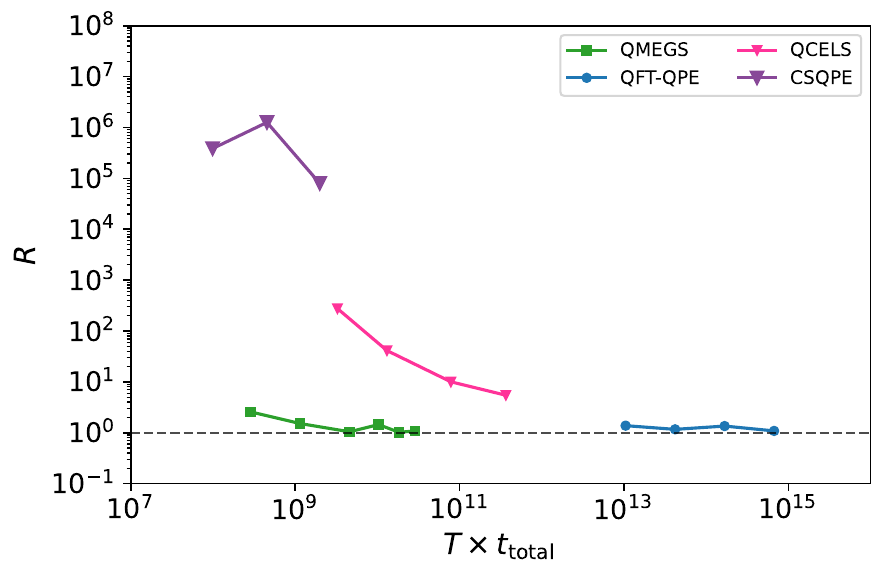}
        \end{minipage}
      } &
      \subfloat[Tail-dense distribution\label{fig:tail_dense}]{
        \begin{minipage}[b]{.32\linewidth}
          \centering
          \includegraphics[width=\linewidth]{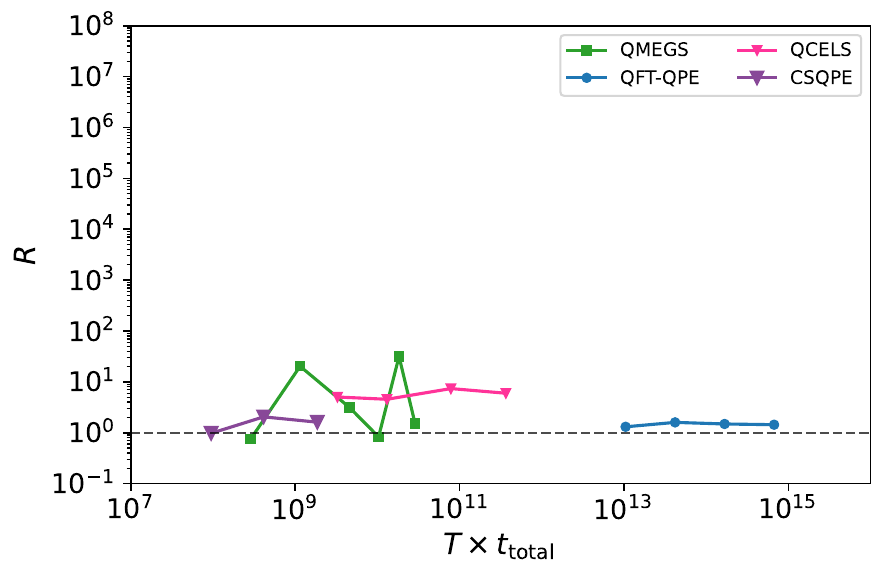}\\[-0.3em]
          \includegraphics[width=\linewidth]{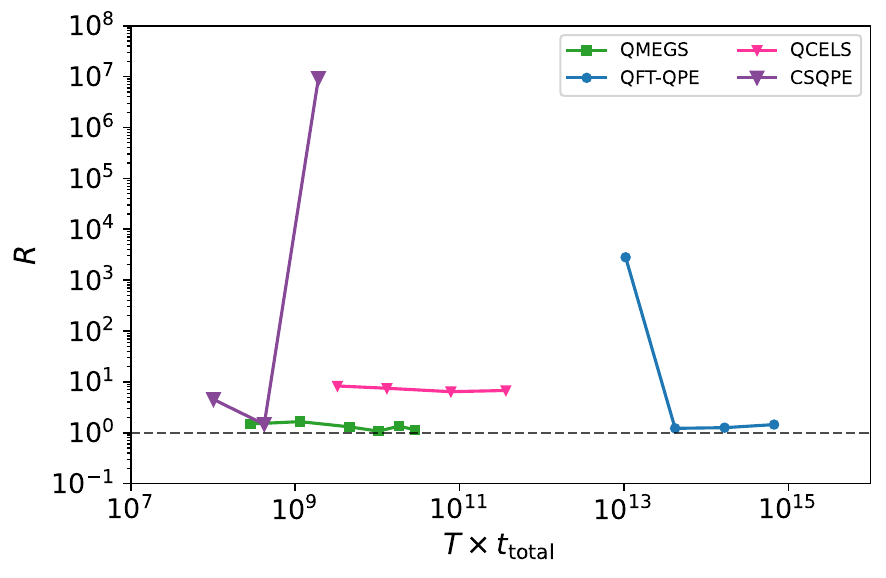}\\[-0.3em]
          \includegraphics[width=\linewidth]{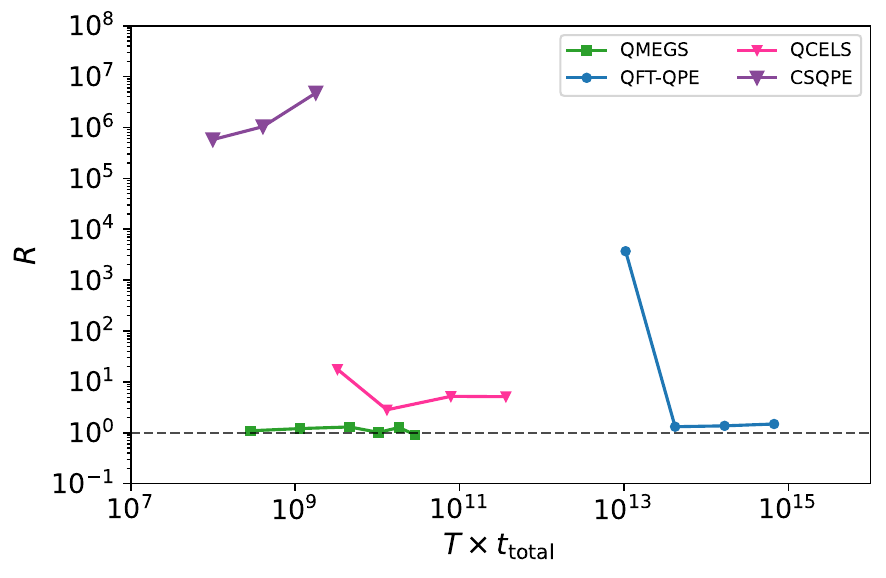}
        \end{minipage}
      } &
      \subfloat[Head-dense distribution\label{fig:head_dense}]{
        \begin{minipage}[b]{.32\linewidth}
          \centering
          \includegraphics[width=\linewidth]{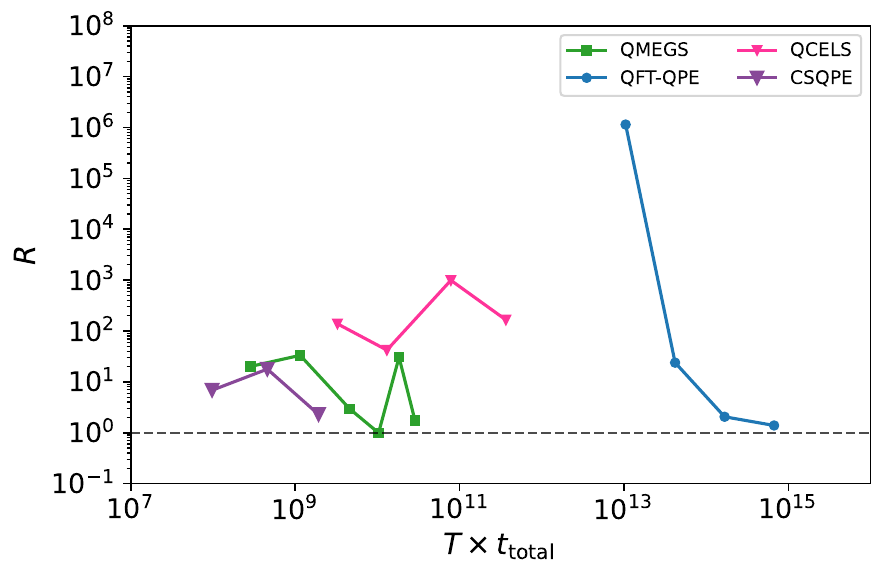}\\[-0.3em]
          \includegraphics[width=\linewidth]{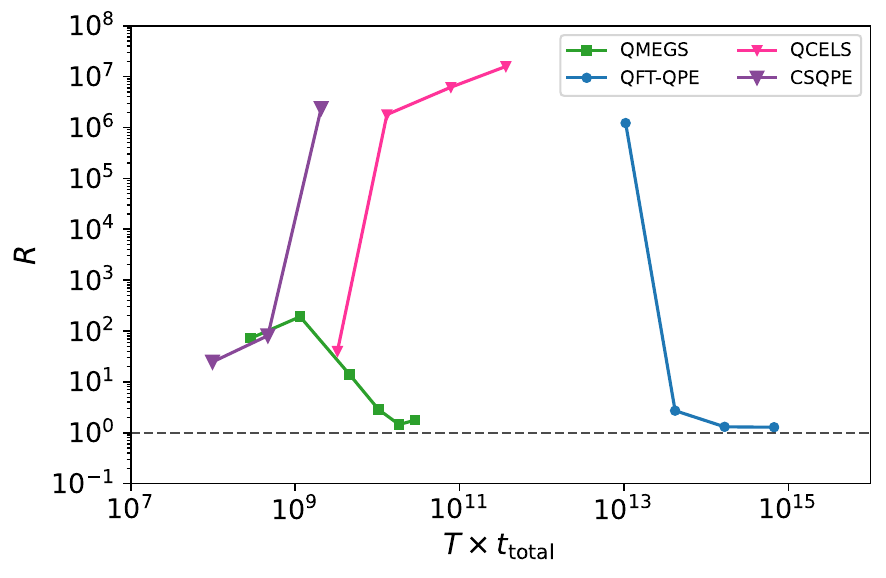}\\[-0.3em]
          \includegraphics[width=\linewidth]{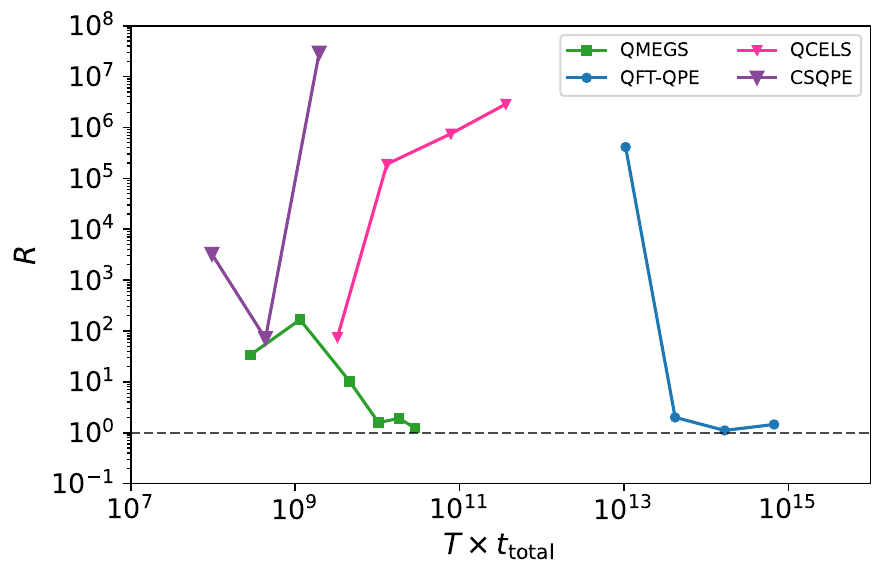}
        \end{minipage}
      }

    \end{tabular}
  \end{adjustbox}
  \caption{Evaluation of the efficiency ratio $R$ at $L=20$. 
(a) Uniform distribution, (b) Tail-dense distribution, and (c) Head-dense distribution. 
The vertical axis represents the ratio $R = T t_{total} \mathrm{MSE}_0 / (\gamma/g_0)$, where the horizontal dashed line at $R=1$ indicates the theoretical lower bound. 
The plots compare the performance of QMEGS, CSQPE, QCELS, and QFT-QPE for overlap parameters $\alpha \in \{0.1, 0.4, 0.7\}$.}
  \label{fig:comp_mse_crlb}
\end{figure*}

\subsection{Comparison of QFT-QPE and HT-QPE}\label{comparison_qpe}
In Secs.~\ref{Theory}, we showed that for both QFT-QPE and HT-QPE, the theoretical lower bound of $Tt_{\text{total}}$ can be quantified by $\gamma/g_i$ under the condition $T\gg \Delta_{\min}^{-1}$. Therefore, we can directly compare these methods by looking at the metric $\gamma/g_i$.  Figure~\ref{fig:gamma_grid} shows the dependence of $\gamma/g_0$ on $c_0$ for the uniform distribution, head-dense distribution and tail-dense distribution, respectively. In the region where $c_0$ roughly above $0.7$, the HT-QPE algorithms have a smaller $\gamma/g_0$, but in the region where $c_0$ roughly below 0.7, QFT-QPE has a smaller $\gamma/g_0$. This result indicates that the choice of algorithm should depend on the distribution of $\boldsymbol{c}$.

This result suggests that for problems where it is easy to prepare an initial state with a large overlap with the target eigenstate, HT-QPE, which requires fewer ancillary qubits, is a practical option. On the other hand, for problems where preparing an initial state with high overlap to a target eigenstate is difficult, QFT-QPE, which is robust even with low overlap, becomes the preferred choice.

\subsection{Performance of practical QPE algorithms relative to the derived lower bound}
\label{subsec:comparison_lower_bound}

In this section, we evaluate the efficiency of practical QPE algorithms by comparing their actual cost products against the theoretical lower bounds derived for an unbiased estimator. Recall from Eq.~\eqref{lower_bound} that the quantity $\gamma/g_i$ represents the theoretical lower bound of the product $T t_{\text{total}} \mathrm{MSE}_i$. To assess how efficiently a given algorithm utilizes the information obtained from the quantum circuit, we calculate the product $T t_{\text{total}} \mathrm{MSE}_i$ using the MSE obtained from numerical simulations and compare it with the theoretical limit $\gamma/g_i$. For clarity in our numerical analysis, we quantify this efficiency using the ratio $R$, defined as:\begin{equation}R = \frac{T t_{\text{total}} \mathrm{MSE}_i}{\gamma/g_i}.\end{equation} An algorithm with $R$ close to 1 indicates that it operates near the fundamental limit of information efficiency. We perform numerical simulations for QMEGS, QCELS, CSQPE, and QFT-QPE to compute their respective MSEs. These simulations are conducted by sampling directly from the theoretical probability distributions derived in Sec.~\ref{Theory}, rather than simulating the full quantum circuits. The settings for the eigenphase distributions and overlaps follow those described in the previous sections. Specifically for the overlap coefficients, we select $\alpha \in \{0.1, 0.4, 0.7\}$ to highlight the crossover regime where the relative advantage of the cost-product lower bounds between QFT-QPE and HT-QPE reverses.

Figure~\ref{fig:comp_mse_crlb} presents the ratio $R$ for the uniform, tail-dense, and head-dense distributions with $\alpha \in \{0.1, 0.4, 0.7\}$ at $L=20$. 
Note that we evaluate $R$ at discrete values of the cost product $Tt_{\text{total}}$ because a continuous sweep across the entire parameter range would entail a prohibitive total computational cost. 
The results show that both QMEGS and QFT-QPE achieve a ratio $R \approx 1$ across the tested regimes, whereas other algorithms exhibit larger deviations.

Regarding QFT-QPE, our results extend previous findings.
While prior studies have confirmed that QFT-QPE achieves the CRLB when the input is a pure eigenstate ($c_0=1$)~\cite{2409.15752}, its behavior for mixed inputs ($c_0 < 1$) remained less explored.
Our numerical results demonstrate that QFT-QPE maintains $R \approx 1$ even in the low-overlap regime, enabled by the algorithmic improvements described in Appendix~\ref{app:curve_fitting_implementation}. Although QFT-QPE exhibits a larger horizontal-axis value, this is because we choose $N_{\text{s}}$ sufficiently large to ensure a good fit to the entire empirical distribution.

For the HT-QPE schemes, QMEGS distinguishes itself by consistently satisfying $R \approx 1$, unlike QCELS and CSQPE.
An $R$ value close to unity indicates that the algorithm utilizes the information obtained from the quantum circuit in a nearly optimal manner, effectively saturating the theoretical lower bound.
In contrast, QCELS and CSQPE show $R$ values significantly larger than $1$.
Since these algorithms share the same fundamental Hadamard-test circuit structure as QMEGS,  the observed gap suggests that the inefficiency lies in the classical post-processing stage; specifically, the classical estimators employed in QCELS and CSQPE fail to fully extract the available information contained in the measurement data, leading to an effective information loss.

Finally, we emphasize that the lower bound given by Eq.~\eqref{lower_bound} is derived under the assumption of unbiased estimators. In practical algorithms, the estimators employed can in general exhibit small biases due to finite-sample effects or implementation details in numerical optimization. A rigorous lower bound for general (possibly biased) estimators, for example based on generalized Cramér--Rao inequalities, is beyond the scope of the present framework and is left for future work. The comparison in this subsection should be interpreted as an evaluation of the distance to the theoretical lower bound under the ``unbiased-estimator assumption'', and the extent to which $T t_{\text{total}} \mathrm{MSE}_i$ approaches $\gamma/g_i$ can be regarded as an indicator of the efficiency of practical algorithms.

\begin{table*}[t]
  \centering
  \caption{Numerical experiment settings in Sec.~IV-E.}
  \label{tab:settings_practical_qpe}
  \renewcommand{\arraystretch}{1.15}
  \begin{tabular}{l l c c}
    \toprule
    Algorithm & $T$ & $N_s$ & $N_t$ \\
    \midrule
    QMEGS
      & $\{250,500,1000,1500,2000,2500\}$ & $2$ & $5000$ \\
    CSQPE
      & $\{200,400,800\}$ & $400$ & --- \\
    QFT-QPE
      & $\{2^{10}-1,2^{11}-1,2^{12}-1,2^{13}-1\}$ & $10^{7}$ & $1$ \\
    ML-QCELS
      & $\{256,512,1024,2048\}$ & $200$ & $500$ \\
    \bottomrule
  \end{tabular}
\end{table*}

\section{Conclusion}\label{conclusion}

In this work, in order to distinguish the performance limits of the quantum circuit structures from the efficiency of classical processing, we established a theoretical framework based on the FIM and the CRLB.
Based on this framework, we derived lower bounds on the cost product $T t_{\text{total}}$ required to achieve a target mean squared error (MSE) for two representative schemes: QFT-QPE and HT-QPE.

Our analysis revealed that the two representative schemes, QFT-QPE and HT-QPE, exhibit distinct scaling behaviors with respect to the overlap $c_{i}$ between the input state and the target eigenstate.
Specifically, we showed that the cost product for QFT-QPE scales as $\Omega(c_{i}^{-1} \text{MSE}_{i}^{-1})$, whereas for HT-QPE it scales as $\Omega(c_{i}^{-2} \text{MSE}_{i}^{-1})$.
This difference implies a performance crossover: HT-QPE offers a lower theoretical bound in the high-overlap regime, while QFT-QPE becomes advantageous when the overlap is small.

We numerically investigated these theoretical findings and the efficiency of practical algorithms.
First, our numerical results explicitly confirmed the crossover phenomenon, showing that the cost-efficiency advantage between the two schemes reverses depending on the initial overlap, as predicted by the theoretical bounds.
Furthermore, we compared the derived limits with the actual costs achieved by practical phase estimation algorithms.
The results demonstrated that for schemes such as QMEGS and curve-fitted QPE, the product $T t_{\text{total}} \text{MSE}_{i}$ closely approaches our derived lower bounds.
This suggests that these algorithms utilize the available statistical information in a nearly optimal manner.

In future work, it will be important to extend this analysis to adaptive phase estimation protocols~\cite{time_adaptive_qpe}.
Additionally, analyzing the lower bounds in the non-asymptotic regime, where the maximum evolution time $T$ is limited, remains a crucial direction to fully understand the practical performance limits of phase estimation.

\section*{Acknowledgement}
This work is supported by MEXT Quantum Leap Flagship Program (MEXT Q-LEAP) Grant No. JPMXS0120319794, and JST COI-NEXT Grant No. JPMJPF2014, JST Moonshot R\&D Grant No. JPMJMS256J.
K.M. is supported by JST FOREST Grant No. JPMJFR232Z and JSPS KAKENHI Grant No. 23H03819.

\bibliographystyle{apsrev4-2}
\bibliography{bib.bib}

\section{Appendix}

\subsection{FIM of QFT-QPE}\label{appendix_qft}
We compute the FIM for $(\boldsymbol{\theta},\boldsymbol{c})$ obtained from a single sample drawn from the probability distribution in Eq.~\eqref{probqft}, can be analytically calculated as follows:
\begin{equation}
  \boldsymbol{\mathcal{I}}^{\mathrm{QFT}}(\boldsymbol{\theta},\boldsymbol{c}|T=2^n-1)
  =
  \begin{bmatrix}
    \mathcal{I}^{\theta\theta} & \mathcal{I}^{\theta c} \\
    \mathcal{I}^{c\theta}     & \mathcal{I}^{cc}
  \end{bmatrix},
\end{equation}
where the components of each block matrix are given for indices $i,j \in \{0,\dots,L-1 \}$:
\begin{align}
\label{FIM_QFT}
\mathcal{I}^{\theta\theta}_{i,j}
&=
\sum_{y=0}^{2^{n}-1}
\frac{%
      \dfrac{c_i c_j}{2^{4n}}\,
      4\,D_{2n}\!\bigl(\phi_{i,y}\bigr) D_{2n}\!\bigl(\phi_{j,y}\bigr)
      D'_{2n}\!\bigl(\phi_{i,y}\bigr) D'_{2n}\!\bigl(\phi_{j,y}\bigr)
     }{%
      \dfrac{1}{2^{2n}}\displaystyle\sum^{L-1}_{l=0} c_l
      D_{2n}^{2}\!\bigl(\phi_{l,y}\bigr)
     }, \\[6pt]
\mathcal{I}^{cc}_{i,j}
&=
\sum_{y=0}^{2^{n}-1}
\frac{1}{2^{2n}}\,
\frac{%
      D_{2^n}^{2}\!\bigl(\phi_{i,y}\bigr)\,
      D_{2^n}^{2}\!\bigl(\phi_{j,y}\bigr)
     }{%
      \displaystyle\sum^{L-1}_{l=0} c_l
      D_{2^n}^{2}\!\bigl(\phi_{l,y}\bigr)
     }, \\[6pt]
\mathcal{I}^{\theta c}_{i,j}
&=
\sum_{y=0}^{2^{n}-1}
\frac{%
      \dfrac{c_i}{2^{2n}}\,
      2\,D_{2^n}\!\bigl(\phi_{i,y}\bigr) D'_{2^n}\!\bigl(\phi_{i,y}\bigr)
      D_{2^n}^{2}\!\bigl(\phi_{j,y}\bigr)
     }{%
      \displaystyle\sum^{L-1}_{l=0} c_l
      D_{2^n}^{2}\!\bigl(\phi_{l,y}\bigr)
     }, \\[6pt]
\mathcal{I}^{c\theta}_{i,j}
&= \mathcal{I}^{\theta c}_{j,i}.
\end{align}

\subsection{Exact evaluation of Fisher Information of QFT-QPE for an Eigenstate}
\label{app:exact_fim}

In this section, we provide an exact derivation of the Fisher information for QFT-QPE when the input state is an eigenstate (i.e., $c_i=1$). 
Let $M = 2^n$ be the dimension of the ancillary Hilbert space. The Dirichlet kernel relevant to our probability distribution is given by:
\begin{equation}
    D_M(x) = \frac{\sin(Mx/2)}{\sin(x/2)}.
\end{equation}
To evaluate the Fisher information, we need to calculate the sum of the squared derivative of this kernel.
Using logarithmic differentiation, the derivative $D_M'(x)$ can be expressed as:
\begin{align}
    \frac{D_M'(x)}{D_M(x)} &= \frac{d}{dx} \left[ \log \sin\left(\frac{Mx}{2}\right) - \log \sin\left(\frac{x}{2}\right) \right] \notag \\
    &= \frac{M}{2} \cot\left(\frac{Mx}{2}\right) - \frac{1}{2} \cot\left(\frac{x}{2}\right).
\end{align}
Thus, we have
\begin{equation}
    D_M'(x) = \frac{D_M(x)}{2} \left( M \cot\left(\frac{Mx}{2}\right) - \cot\left(\frac{x}{2}\right) \right),
\end{equation}
and its square is
\begin{equation}
    \left(D_M'(x)\right)^2 = \frac{D_M(x)^2}{4} \left( M \cot\left(\frac{Mx}{2}\right) - \cot\left(\frac{x}{2}\right) \right)^2.
\end{equation}

In the context of QFT-QPE, the variable $x$ is related to the phase $\theta$ and the measurement outcome $y$ by $x = \theta - \frac{2\pi y}{M}$ for $y=0, 1, \dots, M-1$.
Let us define auxiliary variables:
\begin{equation}
    a_y = \frac{x}{2} = \frac{\theta}{2} - \frac{\pi y}{M}, \quad u = \frac{M\theta}{2}.
\end{equation}
Using these variables, we observe that $\frac{Mx}{2} = u - \pi y$, which implies $\cot(Mx/2) = \cot(u - \pi y) = \cot u$.
Furthermore, the kernel squared can be rewritten as:
\begin{equation}
    D_M(x)^2 = \frac{\sin^2(u - \pi y)}{\sin^2 a_y} = \frac{\sin^2 u}{\sin^2 a_y} = \sin^2 u \csc^2 a_y.
\end{equation}
Substituting these back into the expression for $(D_M'(x))^2$, we obtain:
\begin{equation}
    \left(D_M'(x)\right)^2 = \frac{\sin^2 u}{4} \csc^2 a_y \left( M \cot u - \cot a_y \right)^2.
\end{equation}
Expanding the square term and taking the sum over $y$, we get:
\begin{align}
    \sum_{y=0}^{M-1} (D_M'(x))^2 &= \frac{\sin^2 u}{4} \Bigg[ M^2 \cot^2 u \sum_{y=0}^{M-1} \csc^2 a_y \notag \\
    &\quad - 2M \cot u \sum_{y=0}^{M-1} \csc^2 a_y \cot a_y \notag \\
    &\quad + \sum_{y=0}^{M-1} \csc^2 a_y \cot^2 a_y \Bigg].
\end{align}
To evaluate these sums, we utilize the following trigonometric identities:
\begin{equation}
    \sum_{y=0}^{M-1} \csc^2\left(\frac{\theta}{2} - \frac{\pi y}{M}\right) = M^2 \csc^2\left(\frac{M\theta}{2}\right) = M^2 \csc^2 u,
\end{equation}
\begin{equation}
    \sum_{y=0}^{M-1} \csc^2 a_y \cot a_y = M^3 \csc^2 u \cot u,
\end{equation}
\begin{equation}
    \sum_{y=0}^{M-1} \csc^2 a_y \cot^2 a_y = M^4 \csc^4 u - \frac{1}{3}M^2(1+2M^2) \csc^2 u.
\end{equation}
Substituting these identities into the summation, we find that the terms involving $\cot u$ cancel out significantly:
\begin{align}
    \sum_{y=0}^{M-1} (D_M'(x))^2 &= \frac{\sin^2 u}{4} \left[ \frac{1}{3} M^2 (M^2 - 1) \csc^2 u \right] \notag \\
    &= \frac{1}{12} M^2 (M^2 - 1),
\end{align}
where we used $\sin^2 u \csc^2 u = 1$.

Finally, the Fisher information for QFT-QPE with an eigenstate input is given by:
\begin{equation}
    \mathcal{I}^{\text{QFT}} = \sum_{y=0}^{M-1} \frac{4 (D_M'(x))^2}{M^2},
\end{equation}
Substituting our result:
\begin{equation}
    \mathcal{I}^{\text{QFT}} = \frac{4}{M^2} \cdot \frac{1}{12} M^2 (M^2 - 1) = \frac{M^2 - 1}{3}.
\end{equation}
Recall that $M = 2^n$. Therefore, we arrive at the exact scaling:
\begin{equation}
    \mathcal{I}^{\text{QFT}} = \frac{4^n - 1}{3}.
\end{equation}
This confirms that the Fisher information scales as $\mathcal{O}(2^{2n}) = \mathcal{O}(T^2)$ for an eigenstate.

\subsection{Scaling of the diagonal components of the FIM for QFT-QPE}\label{app:qftqpe_gi_scaling}
In this section, we evaluate the scaling of the diagonal elements of the Fisher Information Matrix (FIM) for QFT-QPE.
Since the FIM has a block structure corresponding to the phase parameters $\theta$ and overlap parameters $c$, we focus on the diagonal component of the $\theta\theta$ block, denoted as $(\mathcal{I}^{\text{QFT}})^{\theta\theta}_{i,i}$.
We explicitly derive both an upper bound and a lower bound to establish that the information acquisition efficiency scales linearly with the overlap $c_i$.
The core of our proof lies in analyzing the behavior of the squared Dirichlet kernel $D_M^2(x)$ and its derivative; specifically, the upper bound is derived from the global summation properties, while the lower bound is rigorously established by extracting the dominant contribution from measurement outcomes in the vicinity of the target eigenphase.

Throughout this analysis, we operate under the following assumptions:
\begin{enumerate}
    \item[(i)] The eigenphases are separated by a $\Delta_{\min} := \min_{j\neq k} |\theta_j - \theta_k| > 0$.
    \item[(ii)] $M$ is sufficiently large to resolve the gap, specifically $\pi/M \le \Delta_{\min}/2$.
    \item[(iii)] The overlap $c_i$ is sufficiently large relative to the system size $M$ to ensure that the target signal dominates the spectral leakage from other eigenstates:
\begin{equation}
    c_i \ge \frac{2\pi^4}{M^2 \Delta_{\min}^2}.
    \label{eq:ci_condition}
\end{equation}
This condition is a sufficient (but not necessary) assumption introduced solely to simplify the bounding of leakage terms; it guarantees that the contribution from the $i$-th eigenstate dominates the cross-talk from other eigenstates in the large-$T$ regime.
\end{enumerate}

Recall the explicit formula for the $\theta\theta$ block element derived in Eq.\eqref{FIM_QFT}:
\begin{equation}
    (\mathcal{I}^{\text{QFT}})^{\theta\theta}_{i,i}
    =\sum_{y=0}^{M-1}
    \frac{4c_i^2\,D_M^2(\phi_{i,y})\,(D_M'(\phi_{i,y}))^2}{M^2\sum_{\ell=0}^{L-1} c_\ell D_M^2(\phi_{\ell,y})},
    \label{eq:Omega_start}
\end{equation}

\subsubsection*{Lower bound: $\Omega(c_i T^2)$}
First, we derive the lower bound. Let $y_0 \in \{0, \dots, M-1\}$ be the index minimizing the distance to $\theta_i$, satisfying $|\phi_{i,y_0}| \le \pi/M$.
For any other eigenphase $\ell \neq i$, the triangle inequality yields:
\begin{equation}
\begin{split}
    |\phi_{\ell,y_0}|
    &= \left| \theta_\ell - \frac{2\pi y_0}{M} \right| \\
    &\ge |\theta_\ell-\theta_i| - |\phi_{i,y_0}| \\
    &\ge \frac{\Delta_{\min}}{2}.
\end{split}
\end{equation}
Using the bound $D_M^2(x) \le \csc^2(x/2)$, the contribution from $\ell \neq i$ to the denominator at $y=y_0$ is bounded by the constant $B_\Delta := \csc^2(\Delta_{\min}/4)$.
Thus, the total denominator satisfies:
\begin{equation}
    \sum_{\ell=0}^{L-1} c_\ell D_M^2(\phi_{\ell,y_0})
    \le c_i D_M^2(\phi_{i,y_0}) + (1-c_i)B_{\Delta}.
\end{equation}
Under the assumption (iii) in Eq.~\eqref{eq:ci_condition}, the signal term $c_i D_M^2(\phi_{i,y_0})$ dominates the noise term. 
Proof: Using Jordan's inequality, we have strict bound $B_\Delta \le 4\pi^2/\Delta_{\min}^2$. Combined with $D_M^2 \ge 4M^2/\pi^2$, the condition $c_i \ge 2 B_\Delta / D_M^2$ (implied by Eq.~\eqref{eq:ci_condition}) guarantees $(1-c_i)B_{\Delta} \le \frac{1}{2} c_i D_M^2(\phi_{i,y_0})$.
Consequently, we obtain the bound for the denominator:
\begin{equation}
    \sum_{\ell=0}^{L-1} c_\ell D_M^2(\phi_{\ell,y_0})
    \le \frac{3}{2} c_i D_M^2(\phi_{i,y_0}).
\end{equation}
Substituting this into Eq.~\eqref{eq:Omega_start} for $y=y_0$, we get:
\begin{equation}
    (\mathcal{I}^{\text{QFT}})^{\theta\theta}_{i,i}
    \ge \frac{8}{3} c_i \frac{(D_M'(\phi_{i,y_0}))^2}{M^2}.
\end{equation}
To ensure the derivative term is non-negligible, we select a neighbor $\bar{y} \in \{y_0-1, y_0+1\}$ such that no cancellation occurs in the logarithmic derivative formula.
For this $\bar{y}$, utilizing the property $|\cot(\phi_{i,\bar{y}}/2)| \ge \sqrt{2}M/\pi$, we derived that the derivative scales as:
\begin{equation}
    \frac{(D_M'(\phi_{i,\bar{y}}))^2}{M^2} \ge \frac{2}{\pi^4} M^2.
\end{equation}
Combining these results, we arrive at the explicit lower bound:
\begin{equation}
\begin{split}
    (\mathcal{I}^{\text{QFT}})^{\theta\theta}_{i,i}
    &\ge \frac{8}{3} c_i \cdot \frac{2}{\pi^4} M^2 \\
    &= \frac{16}{3\pi^4} c_i M^2 \\
    &= \Omega(c_i T^2).
\end{split}
\end{equation}

\subsubsection*{Upper bound: $O(c_i T^2)$}
Next, we establish the upper bound.
We utilize the fact that the denominator is lower-bounded by the single term corresponding to the target eigenstate:
\begin{equation}
    \sum_{\ell=0}^{L-1} c_\ell D_M^2(\phi_{\ell,y}) \ge c_i D_M^2(\phi_{i,y}).
\end{equation}
Substituting this inequality into Eq.~\eqref{eq:Omega_start}, we can bound the entire sum:
\begin{equation}
\begin{split}
    (\mathcal{I}^{\text{QFT}})^{\theta\theta}_{i,i}
    &\le \sum_{y=0}^{M-1} \frac{4c_i^2 D_M^2(\phi_{i,y}) (D_M'(\phi_{i,y}))^2}{M^2 \cdot c_i D_M^2(\phi_{i,y})} \\
    &= \frac{4c_i}{M^2} \sum_{y=0}^{M-1} (D_M'(\phi_{i,y}))^2.
\end{split}
\end{equation}
Using the exact identity for the sum of squared derivatives, $\sum_{y} (D_M'(\phi_{i,y}))^2 = \frac{1}{12} M^2 (M^2-1)$, we obtain:
\begin{equation}
\begin{split}
    (\mathcal{I}^{\text{QFT}})^{\theta\theta}_{i,i}
    &\le \frac{4c_i}{M^2} \cdot \frac{1}{12} M^2 (M^2-1) \\
    &= \frac{1}{3} c_i (M^2-1) \\
    &< \frac{1}{3} c_i M^2.
\end{split}
\end{equation}
Since $M \approx T$, this confirms the upper bound $(\mathcal{I}^{\text{QFT}})^{\theta\theta}_{i,i} = O(c_i T^2)$.
By combining the lower and upper bounds, we conclude that the Fisher information scales as $(\mathcal{I}^{\text{QFT}})^{\theta\theta}_{i,i} = \Theta(c_i T^2)$.
Furthermore, numerical evidence indicates that the actual prefactor is close to the derived upper bound, suggesting $(\mathcal{I}^{\text{QFT}})^{\theta\theta}_{i,i} \approx \frac{1}{3} c_i T^2$.

\subsection{FIM of HT-QPE}\label{appendix_ht}
We compute the FIM obtained from two samples: one from the real-part measurement circuit ($W=I$) and one from the imaginary-part measurement circuit ($W=S^{\dagger}$), both at a given time $t=t_k$. In this situation, the FIM for $(\boldsymbol{\theta},\boldsymbol{c})$, denoted as $\boldsymbol{\mathcal{I}}^{\text{HT}}(\boldsymbol{\theta},\boldsymbol{c} | t_k)$, is as follows:
\begin{equation}
  \boldsymbol{\mathcal{I}}^{\text{HT}}(\boldsymbol{\theta},\boldsymbol{c} | t_k)
  \;=\;
  \begin{bmatrix}
    \mathcal{I}^{\theta\theta} & \mathcal{I}^{\theta c} \\[6pt]
    \mathcal{I}^{c\theta}     & \mathcal{I}^{cc}
  \end{bmatrix},
\end{equation}
where the components of each block matrix are given for indices $i,j \in \{0,\dots,L-1 \}$:
\begin{multline}
\mathcal{I}_{i,j}^{\theta\theta}
=
c_i c_j t^{2}_{k} \times
\\*\left(
  \frac{\sin(t_k\theta_i)\sin(t_k\theta_j)}
       {1-\Bigl(\displaystyle\sum^{L-1}_{l=0} c_l\cos(t_k\theta_l)\Bigr)^{2}}
  +
  \frac{\cos(t_k\theta_i)\cos(t_k\theta_j)}
       {1-\Bigl(\displaystyle\sum^{L-1}_{l=0} c_l\sin(t_k\theta_l)\Bigr)^{2}}
\right),\label{eq:ht_FIM_thetatheta}
\end{multline}
\begin{multline}
\mathcal{I}_{i,j}^{\theta c}
=
-\,c_i t_k \times
\\*\left(
  \frac{\sin(t_k\theta_i)\cos(t_k\theta_j)}
       {1-\Bigl(\displaystyle\sum^{L-1}_{l=0} c_l\cos(t_k\theta_l)\Bigr)^{2}}
  +
  \frac{\cos(t_k\theta_i)\sin(t_k\theta_j)}
       {1-\Bigl(\displaystyle\sum^{L-1}_{l=0} c_l\sin(t_k\theta_l)\Bigr)^{2}}
\right),\label{eq:ht_FIM_thetac}
\end{multline}
\begin{multline}
\mathcal{I}_{i,j}^{cc}
=
  \frac{\cos(t_k\theta_i)\cos(t_k\theta_j)}
       {1-\Bigl(\displaystyle\sum^{L-1}_{l=0} c_l\cos(t_k\theta_l)\Bigr)^{2}}
  +
  \frac{\sin(t_k\theta_i)\sin(t_k\theta_j)}
       {1-\Bigl(\displaystyle\sum^{L-1}_{l=0} c_l\sin(t_k\theta_l)\Bigr)^{2}},\label{eq:ht_FIM_cc}
\end{multline}
\begin{align}
\mathcal{I}_{i,j}^{c\theta}
&= \mathcal{I}_{j,i}^{\theta c}.\label{eq:ht_FIM_ctheta}
\end{align}

\subsection{Bounding the information acquisition efficiency in HT-QPE: $g_i=\Theta(c_i^2)$}
\label{app:htqpe_gi_scaling}

This appendix derives the overlap scaling of the information acquisition efficiency $g_i$ for HT-QPE.
We focus on the diagonal element of the FIM associated with the target eigenphase $\theta_i$.

As shown in Appendix~\ref{appendix_ht}, for a single HT-QPE circuit executed at $t$, the diagonal FIM element for $\theta_i$ is
\begin{equation}
\label{eq:I_single_ht_diag}
(\mathcal{I}^{\text{HT}})^{\theta\theta}_{i,i}(t)
= c_i^{\,2}\, t^{2}\, F_i(t),
\end{equation}
where $F_i(t)$ is defined as
\begin{align}
\label{eq:Fi_def}
F_i(t)
&:= \frac{\sin^2(t\theta_i)}{1-C(t)^2} + \frac{\cos^2(t\theta_i)}{1-S(t)^2},\\
\label{eq:CtSt_def}
C(t) &:= \sum_{l=0}^{L-1} c_l \cos(t\theta_l),\qquad
S(t) := \sum_{l=0}^{L-1} c_l \sin(t\theta_l).
\end{align}
The total diagonal Fisher information is
\begin{equation}
\label{eq:I_total_ht_explicit}
(\mathcal{I}^{\text{HT}}_{\mathrm{total}})^{\theta\theta}_{i,i}
:= \mathbb{E}\!\left[\,N_s\sum_{k=1}^{N_t} \mathcal{I}_i(t_k)\right]
= \mathbb{E}\!\left[\,N_s\sum_{k=1}^{N_t} c_i^{\,2}\, t_k^{2}\, F_i(t_k)\right].
\end{equation}

Next, we use uniform bounds for $F_i(t)$:
\begin{equation}
\label{eq:Fi_bounds}
1 \;\le\; F_i(t) \;\le\; F_i^{\max},
\qquad
F_i^{\max} := 1 + \frac{\theta_i^{2}}{\sum_{l=0}^{L-1} c_l \theta_l^{2}}.
\end{equation}
The upper bound in Eq.~\eqref{eq:Fi_bounds} is proven in Appendix~\ref{app:Fi_upper_bound}.

By applying Eq.~\eqref{eq:Fi_bounds} to Eq.~\eqref{eq:I_total_ht_explicit}, we obtain
\begin{equation}
\label{eq:I_total_sandwich}
N_s c_i^{\,2}\, \mathbb{E}\!\left[\sum_{k=1}^{N_t} t_k^2\right]
\;\le\;
(\mathcal{I}^{\text{HT}}_{\mathrm{total}})^{\theta\theta}_{i,i}
\;\le\;
N_s c_i^{\,2}\, F_i^{\max}\, \mathbb{E}\!\left[\sum_{k=1}^{N_t} t_k^2\right].
\end{equation}

We now summarize how $\mathbb{E}\!\left[\sum_{k=1}^{N_t} t_k^2\right]$ scales for representative time schedules.
The proofs are deferred to Appendix~\ref{app:gamma_chi}.
Define the dimensionless constant
\begin{equation}
\label{eq:chi_def}
\chi
\;:=\;
\frac{1}{N_t T^2}\,\mathbb{E}\!\left[\sum_{k=1}^{N_t} t_k^2\right],
\end{equation}

\begin{equation}
    \mathbb{E}\!\left[\sum_{k=1}^{N_t} t_k^2\right]=\chi\,N_t T^2.
\end{equation}

\begin{figure}[h]
\centering
\graphicspath{{./Figure/}}
\includegraphics[width=0.8\hsize]{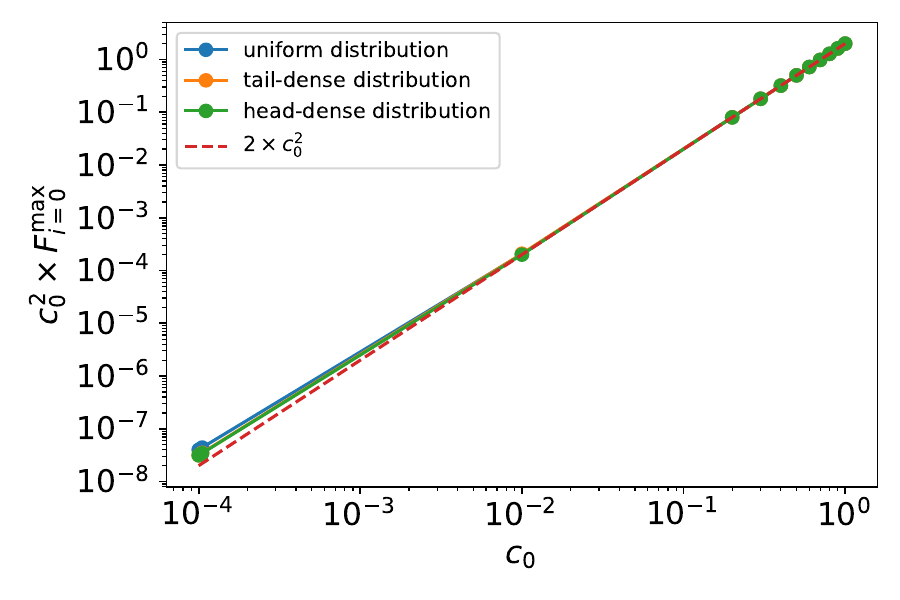}
\caption{Dependence of $c_0^2 F_0^{\max}$ on $c_0$ for $L=10,000$.}
\label{c_i_2_Fi_max_L_10000}
\end{figure}

For example, the following choices yield $\mathbb{E}\!\left[\sum_{k=1}^{N_t} t_k^2\right]=\chi\,N_t T^2$ with protocol-dependent $\chi=O(1)$:

Substituting Eq.~\eqref{eq:chi_def} into Eq.~\eqref{eq:I_total_sandwich}, we obtain the final bound
\begin{equation}
\label{eq:I_total_final_bound}
\chi\, c_i^{\,2}\, N_s N_t T^2
\;\le\;
(\mathcal{I}^{\text{HT}}_{\mathrm{total}})^{\theta\theta}_{i,i}
\;\le\;
F_i^{\max}\,\chi\, c_i^{\,2}\, N_s N_t T^2.
\end{equation}
Recalling the definition of the information acquisition efficiency,
\begin{equation}
\label{eq:gi_def}
g_i := \frac{(\mathcal{I}^{\text{HT}}_{\mathrm{total}})^{\theta\theta}_{i,i}}{N_s N_t T^2},
\end{equation}
Eq.~\eqref{eq:I_total_final_bound} implies
\begin{equation}
\label{eq:gi_bound}
\chi\,c_i^{\,2}\;\le\; g_i \;\le\; F_i^{\max}\,\chi\,c_i^{\,2}.
\end{equation}
Regarding the upper bound, the scaling does not strictly follow $O(c_i^2)$ because $F_i^{\max}$ itself depends on $c_i$. To address this, Figure~\ref{c_i_2_Fi_max_L_10000} shows the dependence of $c_0^2 F_0^{\max}$ on $c_0$ for $L=10,000$ under uniform, tail-dense, and head-dense distributions. The numerical plots indicate that $c_0^2 F_0^{\max}$ remains proportional to $c_0^2$ across all distributions, suggesting that the fluctuations in $F_i^{\max}$ are not significant enough to dominate the overall scaling characteristics. These results lead to the conclusion that the impact of $F_i^{\max}$ on the global scaling is minimal; thus, an evaluation based on $O(c_i^2)$ is sufficiently valid for practical purposes.

\subsection{Derivation of $\gamma$ and $\chi$}
\label{app:gamma_chi}

This appendix evaluates the first and second moments of the time schedule that appear in the total cost and the total Fisher information.
To compute Eqs.~\eqref{def:t_total},~\eqref{eq:fim_total}, it suffices to analyze the following two quantities:
\begin{equation}
\label{eq:two_moments_target}
\mathbb{E}\!\left[\sum_{k=1}^{N_t}|t_k|\right],
\qquad
\mathbb{E}\!\left[\sum_{k=1}^{N_t}t_k^2\right].
\end{equation}
We show that Eq.~\eqref{eq:two_moments_target} can be written as
\begin{equation}
\label{eq:moment_forms_rewrite}
\mathbb{E}\!\left[\sum_{k=1}^{N_t}|t_k|\right]
= \tilde{\gamma} N_t T,
\qquad
\mathbb{E}\!\left[\sum_{k=1}^{N_t}t_k^2\right]
= \chi\,N_t T^2,
\end{equation}
where $\tilde{\gamma}$ satisfies $\gamma = 2\tilde{\gamma}$.

\subsubsection*{QMEGS~\cite{2402.01013}}
In QMEGS, $t$ follows the truncated normal distribution on $[-T,T]$:
\begin{equation}
  a(t)\;=\;\frac1{C\,\sqrt{2\pi}\,T}
         \exp\!\Bigl(-\frac{t^{2}}{2T^{2}}\Bigr)\,
         \mathbf 1_{[-T,T]}(t),
  \qquad
  C\;:=\;\mathrm{erf}\!\Bigl(\tfrac1{\sqrt2}\Bigr).
  \label{eq:truncGauss_app}
\end{equation}
We use the dimensionless variable $x:=t/T\in[-1,1]$.

\medskip
\noindent
\textbf{(i) $\boldsymbol{\tilde{\gamma}_{\mathrm{QMEGS}}}$}
\begin{align}
  \mathbb{E}[|t|]
  &=\int_{-T}^{T}\!|t|\,a(t)\,dt
   \;=\;
     \frac{2}{C\sqrt{2\pi}\,T}
     \int_{0}^{T}\!t\,e^{-t^{2}/(2T^{2})}\,dt \notag\\
  &= \frac{2T}{C\sqrt{2\pi}}
     \int_{0}^{1}\!x\,e^{-x^{2}/2}\,dx
   \;=\;
     \frac{2T}{C\sqrt{2\pi}}
     \Bigl[-e^{-x^{2}/2}\Bigr]_{0}^{1} \notag\\
  &= \frac{T}{C}\sqrt{\frac{2}{\pi}}
     \Bigl(1-\frac1{\sqrt{e}}\Bigr).
\end{align}
Therefore,
\begin{equation}
  \tilde{\gamma}_{\mathrm{QMEGS}}
   \;:=\;\frac{\mathbb{E}[|t|]}{T}
   \;=\;
   \frac{\sqrt{2/\pi}}{C}\,
   \Bigl(1-\frac1{\sqrt{e}}\Bigr).
  \label{eq:gammaQMEGS_app}
\end{equation}

\medskip
\noindent
\textbf{(ii) $\boldsymbol{\chi_{\mathrm{QMEGS}}}$}
\begin{align}
  \mathbb{E}[t^{2}]
  &=\int_{-T}^{T}\!t^{2}\,a(t)\,dt
   \;=\;
     \frac{2}{C\sqrt{2\pi}\,T}
     \int_{0}^{T}\!t^{2}e^{-t^{2}/(2T^{2})}\,dt \notag\\
  &=\frac{2T^{2}}{C\sqrt{2\pi}}
     \int_{0}^{1}\!x^{2}e^{-x^{2}/2}\,dx \notag\\
  &=\frac{2T^{2}}{C\sqrt{2\pi}}
     \Bigl[
       -x\,e^{-x^{2}/2}
       +\sqrt{\frac{\pi}{2}}\,
        \mathrm{erf}\!\Bigl(\tfrac{x}{\sqrt{2}}\Bigr)
     \Bigr]_{0}^{1} \notag\\
  &= T^{2}\Bigl(1-\frac{1}{C}\sqrt{\frac{2}{\pi e}}\Bigr).
\end{align}
Hence,
\begin{equation}
  \chi_{\mathrm{QMEGS}}
  \;:=\;
  \frac{\mathbb{E}[t^{2}]}{T^{2}}
  \;=\;
  1-\frac{1}{C}\sqrt{\frac{2}{\pi e}}.
  \label{eq:chiQMEGS_app}
\end{equation}

\subsubsection*{CSQPE~\cite{CS}}
In CSQPE, $t$ follows the discrete uniform distribution
\begin{equation}
  u(t)\;=\;\frac1{T}, 
  \qquad
  t \in \{1,2,\dots,T\}\subset\mathbb Z_{>0}.
  \label{eq:uniformCS_app}
\end{equation}
Expectation values are evaluated with the discrete variable $k\in\{1,\dots,T\}$.

\medskip
\noindent
\textbf{(i) $\boldsymbol{\tilde{\gamma}_{\mathrm{CS}}}$}
\begin{align}
  \mathbb{E}[|t|]
  &=\frac1{T}\sum_{k=1}^{T} k
   \;=\;
   \frac{T+1}{2}.
\end{align}
Therefore,
\begin{equation}
    \tilde{\gamma}_{\mathrm{CS}}
    \;:=\;
    \frac{\mathbb{E}[|t|]}{T}
    \;=\;
    \frac{T+1}{2T}
    \;\approx\;
    \frac12.
  \label{eq:gammaCS_app}
\end{equation}

\medskip
\noindent
\textbf{(ii) $\boldsymbol{\chi_{\mathrm{CS}}}$}
\begin{align}
  \mathbb{E}[t^{2}]
  &=\frac1{T}\sum_{k=1}^{T} k^{2}
   \;=\;
   \frac{(T+1)(2T+1)}{6}.
\end{align}
Hence,
\begin{equation}
  \chi_{\mathrm{CS}}
  \;:=\;
  \frac{\mathbb{E}[t^{2}]}{T^{2}}
  \;=\;
  \frac{(T+1)(2T+1)}{6T^{2}}
  \;\approx\;
  \frac{1}{3}.
  \label{eq:chiCS_app}
\end{equation}

\subsubsection*{QCELS~\cite{QCELS}}
With a deterministic linear schedule, the times are spaced uniformly with $\Delta t = T/N_t$:
\begin{equation}
\label{eq:linear_schedule_app}
t_k = k\Delta t = k\frac{T}{N_t}, \quad (k = 1, \dots , N_t).
\end{equation}
Since $t_k>0$, we have $|t_k|=t_k$ and thus
\begin{align}
\label{eq:sum_abs_QCELS_app_rewrite}
\sum_{k=1}^{N_t}|t_k|
&= \sum_{k=1}^{N_t} t_k
 = \sum_{k=1}^{N_t} \frac{T}{N_t}k
 = \frac{N_tT}{2}\left(1+\frac{1}{N_t}\right).
\end{align}
Therefore,
\begin{equation}
\label{eq:gamma_QCELS_app}
\tilde{\gamma}_{\mathrm{QCELS}}
:= \frac{1}{N_t T}\sum_{k=1}^{N_t} |t_k|
= \frac{N_t+1}{2N_t}
\;\approx\;\frac12.
\end{equation}
Similarly,
\begin{align}
\label{eq:sum_t2_QCELS_app}
\sum_{k=1}^{N_t} t_k^2
&= \sum_{k=1}^{N_t} \left(\frac{T}{N_t}k\right)^{2}
 = \frac{T^{2}}{N_t^{2}} \sum_{k=1}^{N_t} k^{2}\\
 &= \frac{T^2}{N_t^{2}}\cdot \frac{N_t(N_t+1)(2N_t+1)}{6}.
\end{align}
Hence,
\begin{equation}
\label{eq:chi_QCELS_app}
\chi_{\mathrm{QCELS}}
:= \frac{1}{N_t T^2}\sum_{k=1}^{N_t} t_k^2
= \frac{(N_t+1)(2N_t+1)}{6N_t^{2}}
\;\approx\;\frac13.
\end{equation}

\subsubsection*{Summary}
Table~\ref{tab:gamma-chi} summarizes $\gamma$ and $\chi$ for each schedule.

\begin{table}[h]
  \caption{Comparison of $\gamma$ and $\chi$ for each time-scheduling protocol.}
  \label{tab:gamma-chi}
  \centering
  \renewcommand{\arraystretch}{1.8}
  \begin{tabular}{cccc}
    \toprule
      & \textbf{QMEGS} & \textbf{CSQPE} & \textbf{QCELS} \\
    \midrule
    $\gamma$
      & $\displaystyle
         \frac{2\sqrt{2/\pi}}{C}\!\left(1-\frac{1}{\sqrt{e}}\right)\approx 0.92$
      & $\displaystyle  1$
      & $\displaystyle  1$ \\
    \midrule
    $\chi$
      & $\displaystyle
         1-\frac{1}{C}\sqrt{\frac{2}{\pi e}}\approx 0.29$
      & $\displaystyle  \frac13$
      & $\displaystyle  \frac13$ \\
    \bottomrule
  \end{tabular}
\end{table}


\subsection{Bounding $F_i(t)$ in HT-QPE}
\label{app:Fi_upper_bound}

In this appendix, we prove that the function
\begin{equation}
\label{eq:Fi_def_app}
F_i(t)
:= \frac{\sin^2(t\theta_i)}{1 - C(t)^2} + \frac{\cos^2(t\theta_i)}{1 - S(t)^2},
\end{equation}
is bounded as
\begin{equation}
\label{eq:Fi_bounds_app}
1 \le F_i(t) \le 1 + \frac{\theta_i^2}{\sum_{l=0}^{L-1} c_l \theta_l^2}
=: F_i^{\max},
\end{equation}
where
\begin{equation}
\label{eq:CtSt_def_app}
C(t) := \sum_{l=0}^{L-1} c_l \cos(t\theta_l),
\qquad
S(t) := \sum_{l=0}^{L-1} c_l \sin(t\theta_l).
\end{equation}

\subsubsection{Lower bound}
Since $|C(t)| \le \sum_l c_l |\cos(t\theta_l)| \le \sum_l c_l = 1$, we have $0 \le 1-C(t)^2 \le 1$.
Similarly, $|S(t)| \le 1$ implies $0 \le 1-S(t)^2 \le 1$.
Therefore,
\begin{equation}
F_i(t)
\ge \sin^2(t\theta_i) + \cos^2(t\theta_i) = 1.
\end{equation}

\subsubsection{Upper bound via the singular points}
The only potential obstruction to an upper bound is when the denominators in \eqref{eq:Fi_def_app} approach zero,
namely when $C(t)^2 \to 1$ or $S(t)^2 \to 1$.
We show that $F_i(t)$ remains finite even at such points and converges to the constant $F_i^{\max}$.

\subsubsection{Case 1: $C(t_0)^2 = 1$}
Assume that $C(t_0)^2 = 1$ for some $t_0$.
Since $\sum_l c_l=1$ and each $\cos(t_0\theta_l)\in[-1,1]$,
the equality $|C(t_0)|=1$ implies that all contributing terms must align:
\begin{equation}
\label{eq:cos_alignment}
\cos(t_0\theta_l) = \sigma_C \in \{+1,-1\}
\quad \text{for all } l \text{ with } c_l>0.
\end{equation}
In particular, \eqref{eq:cos_alignment} implies $\sin(t_0\theta_l)=0$ for all such $l$, hence $S(t_0)=0$.

Let $t=t_0+\delta$ with $|\delta|\ll 1$.
Using the Taylor expansion
\(
\cos(t_0\theta_l+\delta\theta_l)
= \cos(t_0\theta_l)\cos(\delta\theta_l)-\sin(t_0\theta_l)\sin(\delta\theta_l)
\)
and $\sin(t_0\theta_l)=0$, we obtain
\begin{align}
C(t)
&= \sum_l c_l \cos(t_0\theta_l)\cos(\delta\theta_l) \notag\\
&= \sigma_C \sum_l c_l \Bigl(1-\frac{\delta^2\theta_l^2}{2}+O(\delta^4)\Bigr) \notag\\
&= \sigma_C\Bigl(1-\frac{\delta^2}{2}\sum_l c_l \theta_l^2 + O(\delta^4)\Bigr).
\label{eq:C_expand}
\end{align}
Therefore,
\begin{equation}
\label{eq:denom_C_expand}
1-C(t)^2
= \delta^2\sum_l c_l\theta_l^2 + O(\delta^4).
\end{equation}

For the numerator, \eqref{eq:cos_alignment} implies $\sin(t_0\theta_i)=0$ and $\cos(t_0\theta_i)=\sigma_C$.
Thus
\begin{equation}
\sin(t\theta_i)=\sin(t_0\theta_i+\delta\theta_i)
= \sigma_C(\delta\theta_i+O(\delta^3)),
\end{equation}
and hence
\begin{equation}
\label{eq:sin2_expand}
\sin^2(t\theta_i)=\theta_i^2\delta^2+O(\delta^4).
\end{equation}
Combining \eqref{eq:denom_C_expand} and \eqref{eq:sin2_expand} yields
\begin{equation}
\label{eq:first_term_limit}
\lim_{t\to t_0}\frac{\sin^2(t\theta_i)}{1-C(t)^2}
=
\frac{\theta_i^2}{\sum_l c_l\theta_l^2}.
\end{equation}

For the second term in \eqref{eq:Fi_def_app}, since $S(t_0)=0$, we have $1-S(t)^2\to 1$ as $t\to t_0$.
Moreover, $\cos^2(t\theta_i)\to \cos^2(t_0\theta_i)=1$.
Therefore,
\begin{equation}
\label{eq:second_term_limit_Ccase}
\lim_{t\to t_0}\frac{\cos^2(t\theta_i)}{1-S(t)^2} = 1.
\end{equation}
From \eqref{eq:first_term_limit} and \eqref{eq:second_term_limit_Ccase},
\begin{equation}
\label{eq:Fi_limit_Ccase}
\lim_{t\to t_0} F_i(t)
= 1 + \frac{\theta_i^2}{\sum_l c_l\theta_l^2}
= F_i^{\max}.
\end{equation}

\subsubsection{Case 2: $S(t_0)^2 = 1$}
The case $S(t_0)^2=1$ is analogous.
The equality $|S(t_0)|=1$ implies alignment
\begin{equation}
\sin(t_0\theta_l)=\sigma_S\in\{+1,-1\}
\quad \text{for all } l \text{ with } c_l>0,
\end{equation}
which in turn yields $C(t_0)=0$.

Let $t=t_0+\delta$.
A Taylor expansion around $t_0$ gives
\begin{equation}
1-S(t)^2
= \delta^2\sum_l c_l\theta_l^2 + O(\delta^4),
\end{equation}
and since $\cos(t_0\theta_i)=0$, we have
\begin{equation}
\cos^2(t\theta_i)=\theta_i^2\delta^2+O(\delta^4).
\end{equation}
Hence,
\begin{equation}
\lim_{t\to t_0}\frac{\cos^2(t\theta_i)}{1-S(t)^2}
=
\frac{\theta_i^2}{\sum_l c_l\theta_l^2},
\qquad
\lim_{t\to t_0}\frac{\sin^2(t\theta_i)}{1-C(t)^2}=1,
\end{equation}
which again yields
\begin{equation}
\label{eq:Fi_limit_Scase}
\lim_{t\to t_0}F_i(t)
= 1 + \frac{\theta_i^2}{\sum_l c_l\theta_l^2}
= F_i^{\max}.
\end{equation}


\subsection{Implementation Details of Curve-fitted QPE for General Input States}
\label{app:curve_fitting_implementation}

In this study, we employ an extended version of the curve-fitted QPE algorithm~\cite{2409.15752} to estimate eigenphases from general input states that may contain multiple dominant components. Unlike the standard approach which typically assumes a single eigenstate, our implementation dynamically determines the model order and employs a multi-start optimization strategy to ensure robustness against local minima.

The estimation procedure begins by executing the quantum circuit shown in Fig.~\ref{fig:QFTQPE_kairo} for $N_s$ shots to obtain the empirical probability distribution, denoted as $\hat{p}_y$, for the measurement outcomes $y \in \{0, \dots, 2^n-1\}$. To adapt the model to the spectral richness of the input state, we first perform peak detection on $\hat{p}_y$. The number of detected peaks, $K$, is explicitly adopted as the number of basis functions for the curve fitting. Based on this, we define the fitting function $f(y; \boldsymbol{\theta}, \boldsymbol{A})$ as a superposition of $K$ squared Dirichlet kernels:
\begin{equation}
    f(y; \boldsymbol{\theta}, \boldsymbol{A}) = \sum_{m=1}^{K} A_m \mathcal{D}_{2^n}^2\left(\theta_m - \frac{2\pi y}{2^n}\right),
\end{equation}
where $\boldsymbol{\theta} = \{\theta_1, \dots, \theta_K\}$ represents the target eigenphases, $\boldsymbol{A} = \{A_1, \dots, A_K\}$ represents the corresponding amplitudes. The parameters are estimated by minimizing the sum of squared residuals between the model $f(y)$ and the empirical data $\hat{p}_y$.

\subsection{Fisher Information and Cost Scaling in Robust Phase Estimation (RPE)}\label{app:RPE}

In this Appendix, we discuss whether the cost lower bound framework derived in this paper is applicable to Robust Phase Estimation (RPE)~\cite{RPE}, a representative phase estimation protocol. We conclude that since the number of sampling steps $N_t$ in RPE depends on the maximum evolution time $T$, it does not satisfy the relation $t_{\text{total}} = \gamma N_t N_s T$ assumed in our work. Consequently, the cost lower bound in Eq.~\eqref{lower_bound} cannot be directly applied. However, we can analytically show that the diagonal components of the FIM itself exhibit the scaling $(\mathcal{I}_{\text{total}}^{\text{HT}})_{i,i}^{\theta\theta} = \Theta(c_i^2 T t_{\text{total}})$. In the following, we evaluate the diagonal elements of the FIM based on the RPE schedule, and then explain why the diagonal approximation used in our paper breaks down for RPE.

First, we define $T$ and $t_{\text{total}}$ to evaluate the diagonal elements of the FIM. RPE employs a protocol that exponentially increases the evolution time $t$, selecting the time sequence as $t_k \in \{2^0, 2^1, \dots, 2^{m-1}\}$. For this schedule, the maximum execution time $T$ and the total execution time $t_{\text{total}}$ are given by
\begin{align}
T &= 2^{m-1}, \\
t_{\text{total}} &= 2N_s \sum^{m-1}_{k=0} 2^k = 2N_s(2^m - 1) = 2N_s(2T - 1).
\end{align}
Next, we evaluate the diagonal elements of the FIM. Since the sampling at each $t_k$ is independent, the diagonal element of the total FIM depends on the sum of squared sampling times $\sum_{k=0}^{m-1} t_k^2 = \sum_{k=0}^{m-1} 4^k = \frac{4^m - 1}{3}$, which satisfies the following inequality:
\begin{equation}
N_s c_i^2 \frac{4^m - 1}{3}
\le
(\mathcal{I}_{\text{total}}^{\text{HT}})_{i,i}^{\theta\theta}
\le
N_s c_i^2 F_i^{\max} \frac{4^m - 1}{3}.
\end{equation}
By substituting $4^m = 4T^2$ and $N_s = \frac{t_{\text{total}}}{2(2T - 1)}$ into the lower bound side and rearranging the terms, we obtain
\begin{equation}
N_s c_i^2 \frac{4^m - 1}{3} = \frac{c_i^2}{3} \frac{t_{\text{total}}}{2(2T - 1)} (4T^2 - 1) = \frac{c_i^2}{3} \frac{2T + 1}{2} t_{\text{total}}.
\end{equation}
Therefore, in the asymptotic regime where $T \gg 1$, the diagonal elements of the FIM are proportional to the product of the total execution time and the maximum execution time, yielding $(\mathcal{I}_{\text{total}}^{\text{HT}})_{i,i}^{\theta\theta} = \Theta(c_i^2 T t_{\text{total}})$.

On the other hand, the off-diagonal elements of the FIM in RPE do not become sufficiently small compared to the diagonal elements, meaning the diagonal approximation $(\mathcal{I}_{\text{total}}^{-1})_{i,i} \approx (\mathcal{I}_{\text{total}})_{i,i}^{-1}$ assumed in this paper does not hold. In other HT-QPE methods discussed in our paper (such as QMEGS and QCELS), the circuit is executed at many different times $t$ ($N_t \gg 1$). During the summation over $t$, the oscillating terms in the off-diagonal elements of the FIM cancel each other out, making the matrix diagonally dominant. In contrast, RPE uses a logarithmic number of steps $N_t = m \approx \log_2 T + 1$, meaning the time sequence is extremely sparse. As a result, the cancellation of the off-diagonal elements does not function sufficiently. If we approximate the Cram\'er-Rao lower bound using only the diagonal elements in this case, it leads to an underestimation of the true lower bound of the estimation error.

Although the unfavorable scaling of robust multiple phase estimation (RMPE)~\cite{RMPE} does not directly follow from the CRLB, it is qualitatively consistent with our observation that RPE-type sampling schedules do not sufficiently suppress off-diagonal Fisher-information terms. This suggests that separating multiple eigenvalue contributions is intrinsically more difficult in this setting than in methods such as QMEGS.


\subsection{Scaling of the diagonal components of the FIM in HT-QPE}\label{scaling_ht}

\begin{figure*}[t]
  \centering
  \begin{adjustbox}{max width=\textwidth,max totalheight=\textheight}
    \begin{tabular}{ccc}
      \subfloat[Uniform distribution\label{fig:uniform_dist}]{
        \begin{minipage}[b]{.32\linewidth}
          \centering
          \includegraphics[width=\linewidth]{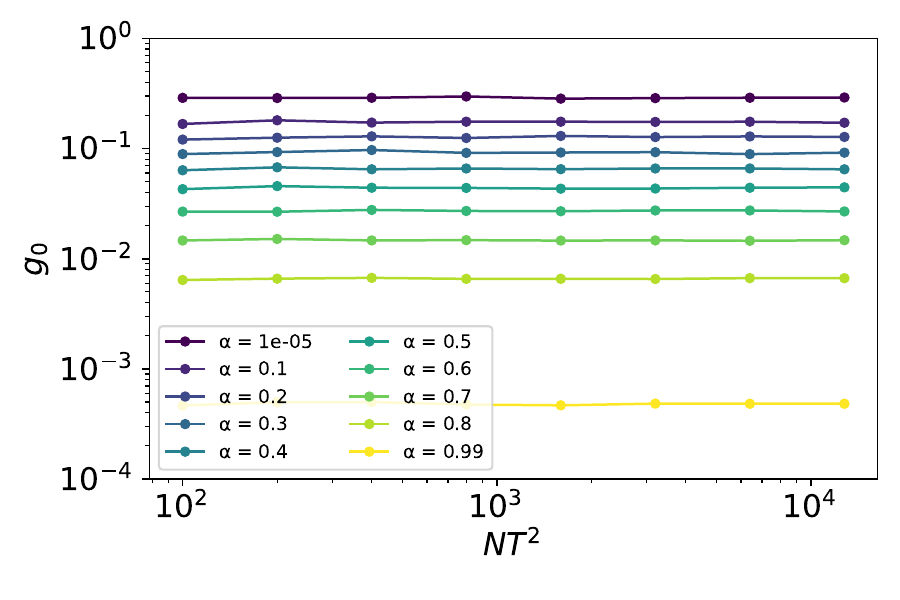}\\[-0.3em]
          \includegraphics[width=\linewidth]{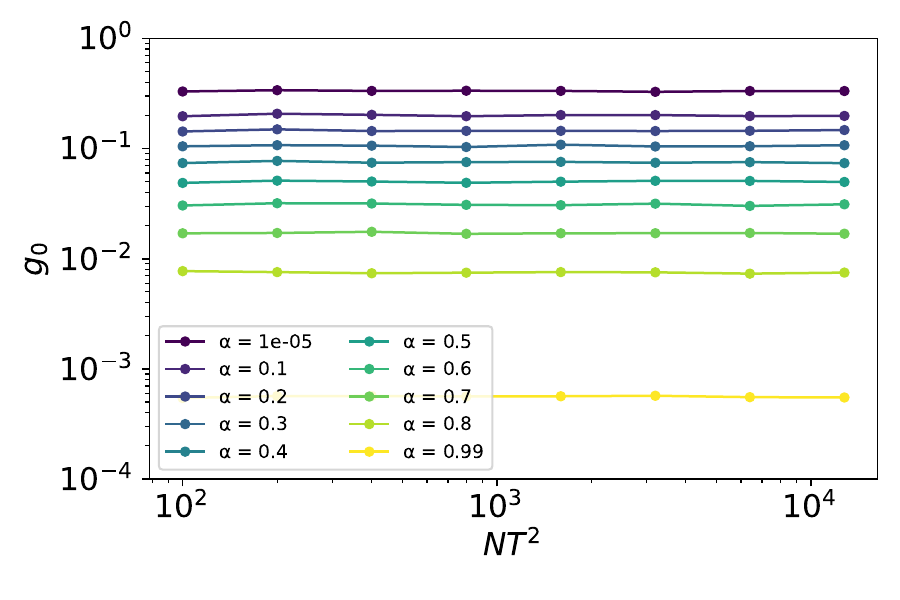}\\[-0.3em]
          \includegraphics[width=\linewidth]{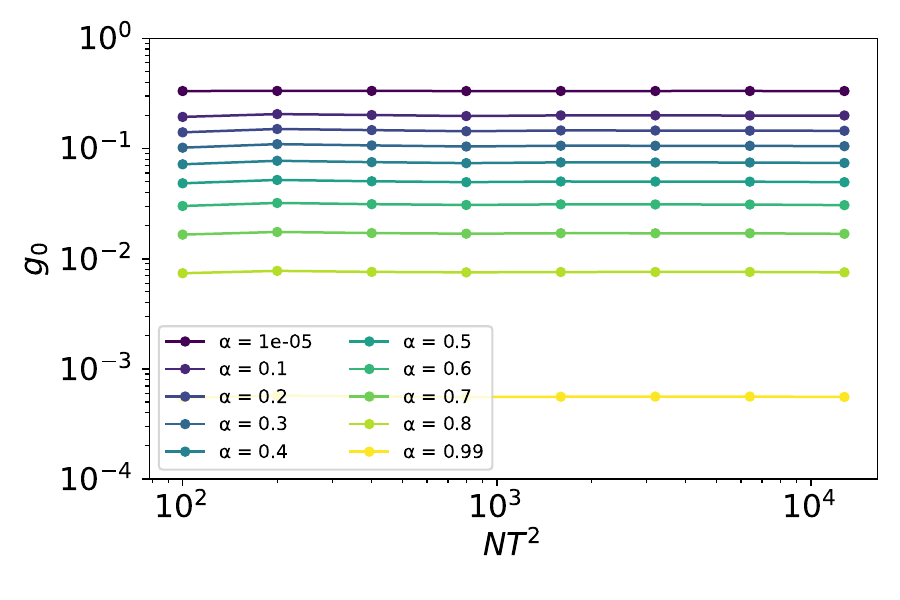}
        \end{minipage}
      } &

      \subfloat[Tail-dense distribution\label{fig:tail_dense}]{
        \begin{minipage}[b]{.32\linewidth}
          \centering
          \includegraphics[width=\linewidth]{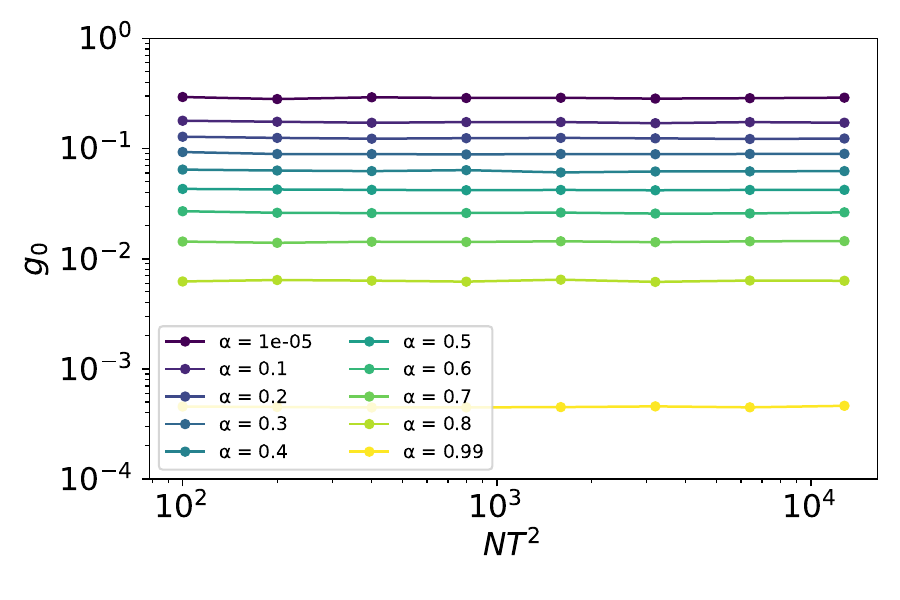}\\[-0.3em]
          \includegraphics[width=\linewidth]{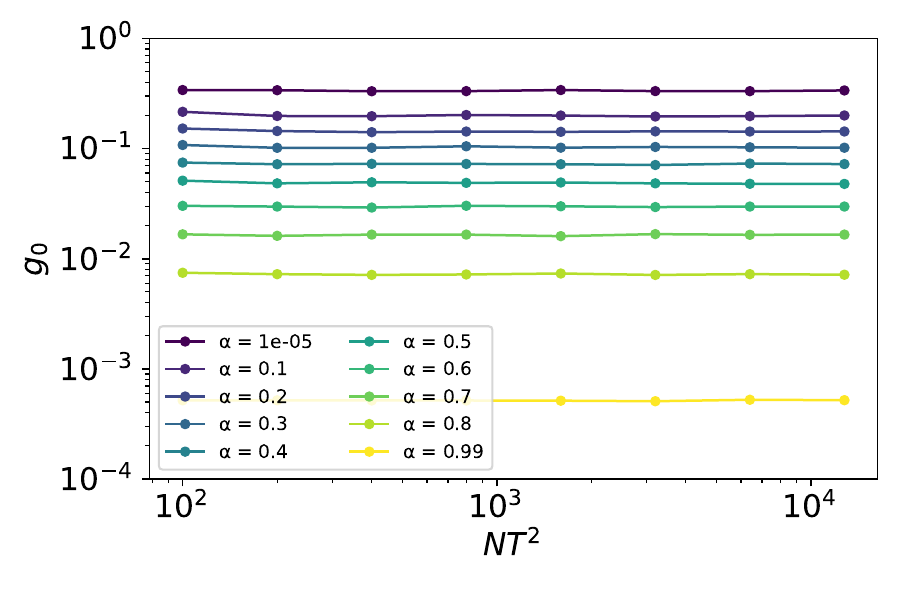}\\[-0.3em]
          \includegraphics[width=\linewidth]{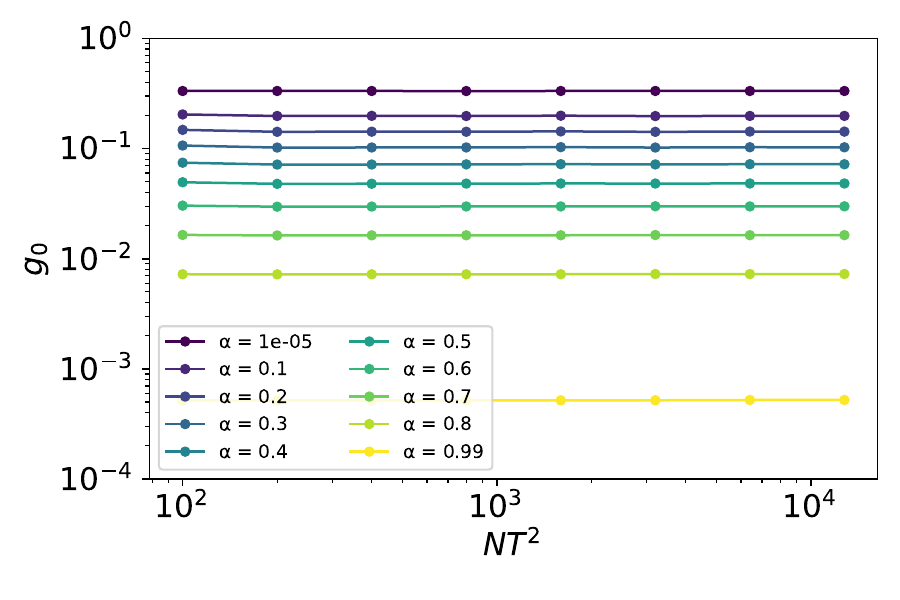}
        \end{minipage}
      } &
      \subfloat[Head-dense distribution\label{fig:head_dense}]{
        \begin{minipage}[b]{.32\linewidth}
          \centering
          \includegraphics[width=\linewidth]{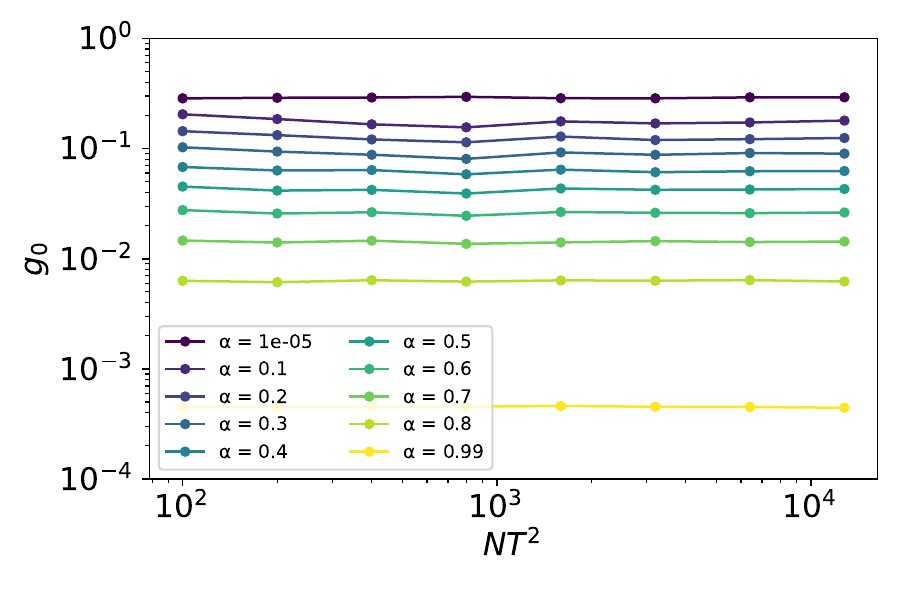}\\[-0.3em]
          \includegraphics[width=\linewidth]{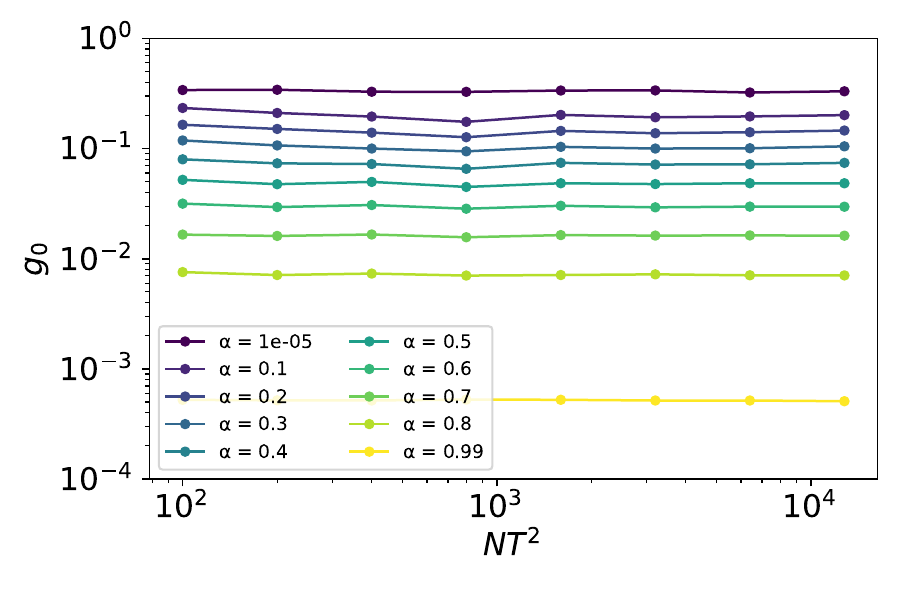}\\[-0.3em]
          \includegraphics[width=\linewidth]{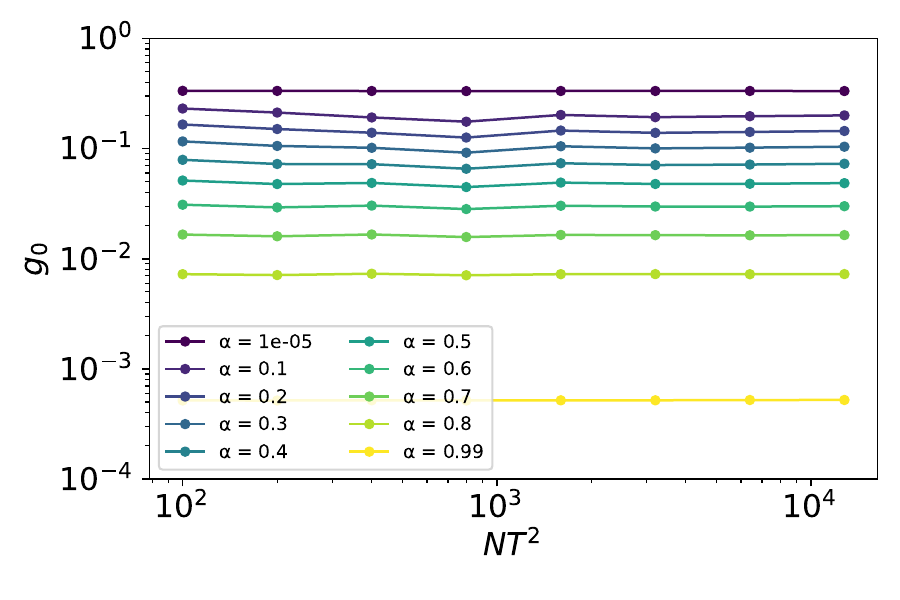}
        \end{minipage}
      } 
    \end{tabular}
  \end{adjustbox}
  \caption{Dependence of $g_0$ on $NT^2$ at $L=20$. (a) Uniform distribution, (b) Tail-dense distribution, and (c) Head-dense distribution. In each subfigure the three stacked panels correspond to QMEGS (top), CSQPE (middle), and QCELS (bottom).}
  \label{fig:diag_fim_grid_order}
\end{figure*}

In this section, we show that the diagonal components $(\mathcal{I}^{\mathrm{HT}}_{\mathrm{total}})_{i,i}$ scale proportionally to $N T^{2}$. 
Figure~\ref{fig:diag_fim_grid_order} show the dependence of $g_0$ on $NT^2$ for the uniform distribution, head-dense distribution, and tail-dense distribution, respectively. All graphs show that $g_0$ is nearly constant with respect to $NT^2$ across all $\alpha$. Therefore, we argue that
\begin{equation}
    (\mathcal{I}^{\text{HT}}_{\text{total}})_{i,i}=g_iNT^2.
\end{equation}

\begin{figure*}[t]
  \centering
  \begin{adjustbox}{max width=\textwidth,max totalheight=\textheight}
    \begin{tabular}{ccc}
      \subfloat[Uniform distribution\label{fig:uniform_dist}]{
        \begin{minipage}[b]{.32\linewidth}
          \centering
          \includegraphics[width=\linewidth]{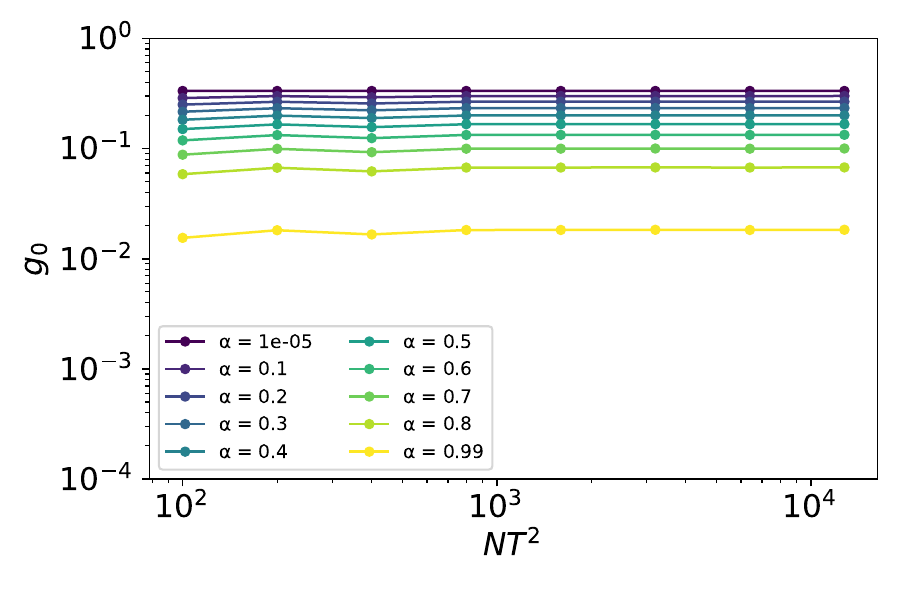}\\[-0.3em]
        \end{minipage}
      } &
      \subfloat[Tail-dense distribution\label{fig:tail_dense}]{
        \begin{minipage}[b]{.32\linewidth}
          \centering
          \includegraphics[width=\linewidth]{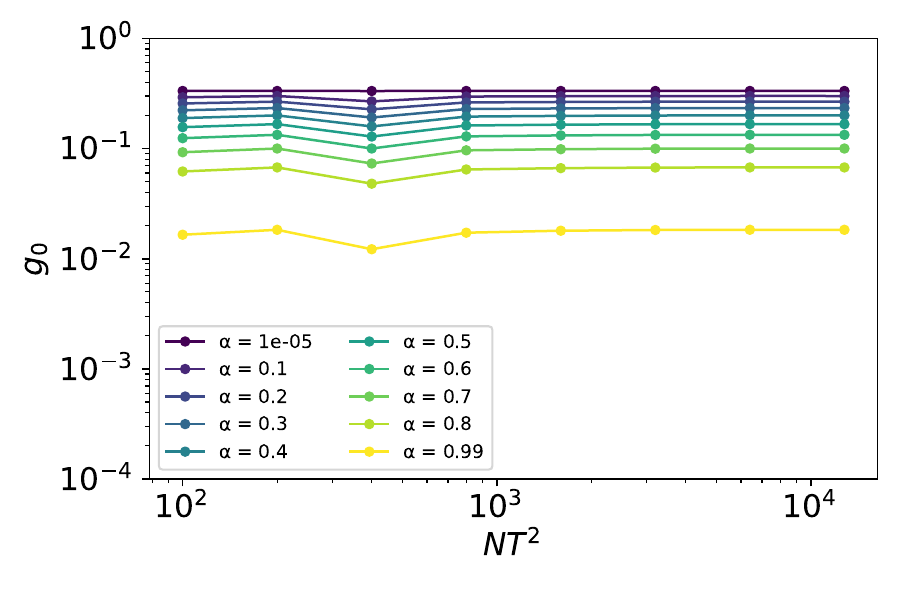}\\[-0.3em]
        \end{minipage}
      } &
      \subfloat[Head-dense distribution\label{fig:head_dense}]{
        \begin{minipage}[b]{.32\linewidth}
          \centering
          \includegraphics[width=\linewidth]{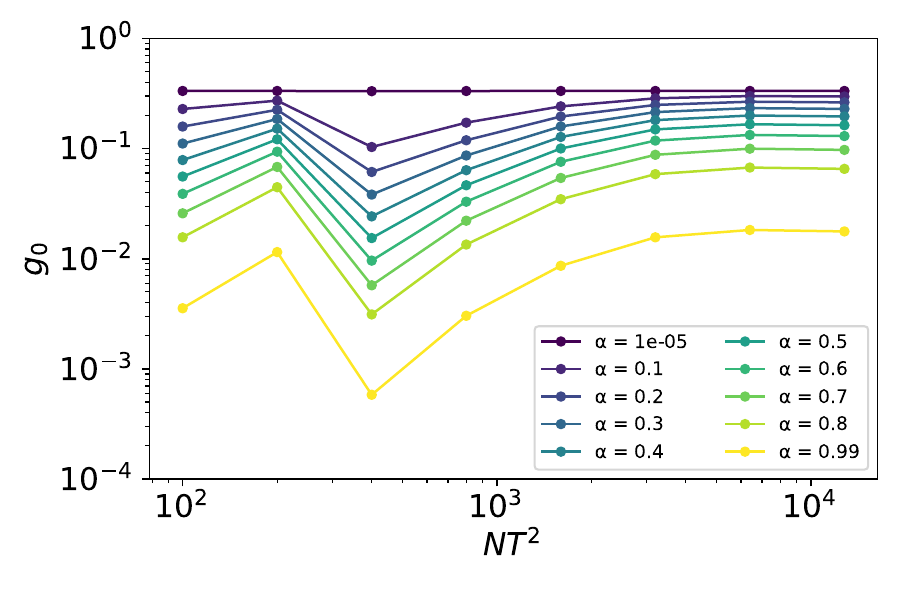}\\[-0.3em]
        \end{minipage}
      }
    \end{tabular}
  \end{adjustbox}
  \caption{Dependence of $g_0$ on $NT^2$ at $L=20$ for QFT-QPE. (a) Uniform distribution, (b) Tail-dense distribution, and (c) Head-dense distribution. }
  \label{fig:diag_fim_grid_order_qft}
\end{figure*}

\end{document}